\documentclass[superscriptaddress,preprint]{revtex4}%
\usepackage{amssymb}
\usepackage{amsmath}
\usepackage{amsfonts}
\usepackage{graphicx}
\usepackage{hyperref}%
\begin{document}
\title{High Energy Scattering Amplitudes of Superstring Theory}
\author{Chuan-Tsung Chan}
\email{ctchan@phys.cts.nthu.edu.tw}
\affiliation{Physics Division, National Center for Theoretical Sciences, Hsinchu, Taiwan, R.O.C.}
\author{Jen-Chi Lee}
\email{jcclee@cc.nctu.edu.tw}
\affiliation{Department of Electrophysics, National Chiao-Tung University, Hsinchu, Taiwan, R.O.C.}
\author{Yi Yang}
\email{yiyang@mail.nctu.edu.tw}
\affiliation{Department of Electrophysics, National Chiao-Tung University, Hsinchu, Taiwan, R.O.C.}
\date{\today }

\begin{abstract}
We use three different methods to calculate the proportionality constants
among high-energy scattering amplitudes of different string states with
polarizations on the scattering plane. These are the decoupling of high-energy
zero-norm states (HZNS), the Virasoro constraints and the saddle-point
calculation. These calculations are performed at arbitrary but fixed mass
level for the NS sector of 10D open superstring. All three methods give the
consistent results, which generalize the previous works on the high-energy 26D
open bosonic string theory. In addition, we discover new leading order
high-energy scattering amplitudes, which are still proportional to the
previous ones, with polarizations \textit{orthogonal} to the scattering plane.
These scattering amplitudes are of subleading order in energy for the case of
26D open bosonic string theory. The existence of these new high-energy
scattering amplitudes is due to the worldsheet fermion exchange in the
correlation functions and is, presumably, related to the high-energy massive
spacetime fermionic scattering amplitudes in the R-sector of the theory.

\end{abstract}
\maketitle


\section{Introduction}

It has long been believed that string theory consists of a huge hidden
symmetry. This is strongly suggested by the ultraviolet finiteness of quantum
string theory, which contains no free parameter and an infinite number of
states. To probe the structure and the origin of the symmetry has been one of
the fundamental issue ever since the discovery of string theory.

The first key idea to uncover the hidden stringy symmetry was to study the
high-energy behavior of the theory, as suggested by Gross in 1988
\cite{Gross}. This was based on the saddle-point calculations of high-energy
fixed-angle string scattering amplitudes in \cite{GM,GrossManes}. There are
two main conjectures of Gross's pioneer work on this subject. The first one is
the existence of an infinite number of linear relations among the scattering
amplitudes of different string states that are valid order by order in
perturbation theory at high energies. The second is that this symmetry is so
powerful as to determine the scattering amplitudes of all the infinite number
of string states in terms of, say, the tachyon scattering amplitudes (for the
bosonic open string case). However, the symmetry charges of his proposed
stringy symmetries were not understood and the proportionality constants among
high-energy scattering amplitudes of different string states were not calculated.

The second key idea to uncover the fundamental symmetry of string theory was
the identification of symmetry charges from an infinite number of stringy
zero-norm states with arbitrarily high spins in the old covariant first
quantized (OCFQ) string spectrum \cite{Lee}. The importance of zero-norm
states and their implication on stringy symmetries were first pointed out
\cite{Lee} in the context of massive $\sigma$-model approach \cite{Labas, Gan}
of string theory. Some implications of the corresponding stringy Ward
identities on the scattering amplitudes were discussed in \cite{JCLee,
MassHeter}. In addition to the continuous symmetries, the discrete T-duality
symmetry was shown to be related to the existence of compactified closed
string soliton zero-norm states \cite{Soliton}. The enhanced gauge symmetry of
N coincident D-branes can also be shown to be related to the existence of
compatified open string zero-norm states at some discrete values of
compatified radii \cite{D-brane}. On the other hand, zero-norm states were
also shown \cite{ChungLee} to carry the spacetime $\omega_{\infty}$ symmetry
\cite{Winfinity} charges of 2D string theory \cite{2Dstring}. This is in
parallel with the work of \cite{Ring} where the ground ring structure of ghost
number zero operators was identified in the BRST quantization. All the above
interesting results of 26D and 2D string theories strongly suggest that a
clearer understanding of zero-norm states holds promise to uncover the
fundamental symmetry of string theory. Incidentally, it was also shown
\cite{KaoLee,CLYang} that off-shell gauge transformations of Witten string
field theory \cite{Witten}, after imposing the no-ghost condition, are
identical to the on-shell stringy gauge symmetries generated by two types of
zero-norm states in the massive $\sigma$-model approach of string theory
\cite{Lee}. Other approaches of stringy symmetries can be found in
\cite{Moore, West, Veneziano, Kubota, Bianchi, Regge, Hagedorn, WS}.

Recently high-energy Ward identities derived from the decoupling of zero-norm
states, which combines the previous two key ideas of probing stringy symmetry,
were used to explicitly prove Gross's two conjectures \cite{ChanLee1,ChanLee2}%
. An infinite number of linear relations among high-energy scattering
amplitudes of different string states were derived. Moreover, these linear
relations can be used to fix the proportionality constants among high-energy
scattering amplitudes of different string states algebraically up to mass
level $M^{2}=6$. Thus there is only one independent component of high-energy
scattering amplitude at each fixed mass level. It is important to discover
that the result of saddle-point calculation in \cite{GM, Gross, GrossManes}
was inconsistent with high-energy stringy Ward identities of zero-norm state
calculation in \cite{ChanLee1,ChanLee2, CHL}. A corrected saddle-point
calculation was given in \cite{CHL}, where the missing terms of the
calculation in \cite{GM, Gross, GrossManes} were identified to recover the
stringy Ward identities. Soon after, the calculations of the proportionality
constants among high-energy scattering amplitudes of different string states
were generalized to arbitrary but fixed mass level \cite{CHLTY1,CHLTY2,CHLTY3}%
. Based on the general formula for the independent component of high-energy
scattering amplitude at each fixed mass level calculated previously in
\cite{ChanLee1,CHL}, one can then derive the general formula of high-energy
scattering amplitude for four arbitrary string states, and express them in
terms of those of tachyons. This completes the general proofs of Gross's two
conjectures on high-energy symmetry of string theory stated above.

In this paper, we consider the high-energy scattering amplitudes for the NS
sector of 10D open superstring theory. Based on the calculations of 26D
bosonic open string \cite{CHLTY1,CHLTY2,CHLTY3}, all the three independent
calculations of bosonic string, namely the decoupling of high-energy zero-norm
states (HZNS), the Virasoro constraints and the saddle-point calculation, can
be generalized to scattering amplitudes of string states with polarizations on
the scattering plane of superstring. All three methods give the consistent
results. In addition, we discover new leading order high-energy scattering
amplitudes, which are still proportional to the previous ones, with
polarizations \textit{orthogonal} to the scattering plane. These scattering
amplitudes are of subleading order in energy for the case of 26D open bosonic
string theory. The existence of these new high-energy scattering amplitudes is
due to the worldsheet fermion exchange in the correlation functions and is,
presumably, related to the high-energy massive fermionic scattering amplitudes
in the R-sector of the theory. We thus conjecture that the validity of Gross's
two conjectures on high-energy stringy symmetry persists for superstring
theory. This paper is organized as follows. In section II, after a brief
review of previous calculations of the decoupling of HZNS for the bosonic
string, we show that the calculations can be generalized to the superstring.
The ratios among the scattering amplitudes of different string states
\textit{with polarizations on the scattering plane} can be determined by using
two types of HZNS in the NS sector. In section III, we use the "dual method",
the Virasoro constraints, to calculate the ratios among the scattering
amplitudes of different string states. The results are consistent with those
obtained in section II. In section IV, a set of scattering amplitudes are
calculated by the saddle-point method to justify our results in sections II
and III. In section V, we present some new high-energy scattering amplitudes
of string states with polarizations \textit{orthogonal} to the scattering
plane. Finally a brief conclusion is given in Section VI.

\section{Decoupling of HZNS}

We will consider four-point correlation functions in this paper. We begin with
a brief review of the high-energy calculation of 26D bosonic open string
theory. At a fixed mass level $M^{2}=2(n-1)$, it was shown that
\cite{CHLTY1,CHLTY2,CHLTY3} in the high-energy limit, only states of the
following form
\begin{equation}
\left\vert n,2m,q\right\rangle \equiv(\alpha_{-1}^{T})^{n-2m-2q}(\alpha
_{-1}^{L})^{2m}(\alpha_{-2}^{L})^{q}\left\vert 0,k\right\rangle ,\text{ where
}n-2m-2q,m,q\geqslant0, \label{1}%
\end{equation}
are relevant for four-point functions. ( we use the notation of \cite{GSW}).
The state in Eq.(\ref{1}) is arbitrarily chosen to be the second vertex of the
four-point function. The other three points can be any string states. We have
defined the normalized polarization vectors of the second string state to be
\cite{ChanLee1,ChanLee2}
\begin{equation}
e^{P}=\frac{1}{M_{2}}(E_{2},\mathrm{k}_{2},0)=\frac{k_{2}}{M_{2}}, \label{2}%
\end{equation}%
\begin{equation}
e^{L}=\frac{1}{M_{2}}(\mathrm{k}_{2},E_{2},0), \label{3}%
\end{equation}%
\begin{equation}
e^{T}=(0,0,1), \label{4}%
\end{equation}
in the CM frame contained in the plane of scattering. In the OCFQ spectrum of
open bosonic string theory, the solutions of physical states conditions
include positive-norm propagating states and two types of zero-norm states.
The latter are \cite{GSW}%
\begin{equation}
\text{Type I}:L_{-1}\left\vert x\right\rangle ,\text{ where }L_{1}\left\vert
x\right\rangle =L_{2}\left\vert x\right\rangle =0,\text{ }L_{0}\left\vert
x\right\rangle =0; \label{5}%
\end{equation}%
\begin{equation}
\text{Type II}:(L_{-2}+\frac{3}{2}L_{-1}^{2})\left\vert \widetilde
{x}\right\rangle ,\text{ where }L_{1}\left\vert \widetilde{x}\right\rangle
=L_{2}\left\vert \widetilde{x}\right\rangle =0,\text{ }(L_{0}+1)\left\vert
\widetilde{x}\right\rangle =0. \label{6}%
\end{equation}
While Type I states have zero-norm at any space-time dimension, Type II states
have zero-norm \emph{only} at D=26. The decoupling of the following Type I
HZNS%
\begin{equation}
L_{-1}\left\vert n-1,2m-1,q\right\rangle =M\left\vert n,2m,q\right\rangle
+(2m-1)\left\vert n,2m-2,q+1\right\rangle \label{7}%
\end{equation}
gives the first high-energy Ward identities%
\begin{equation}
\mathcal{T}^{(n,2m,q)}=(-\frac{2m-1}{M})....(-\frac{3}{M})(-\frac{1}%
{M})\mathcal{T}^{(n,0,q+m)}. \label{8}%
\end{equation}
where $\mathcal{T}^{(n,2m,q)}$ represents the four-point functions with the
second particle at level $n.$ Similarly, the decoupling of the following Type
II HZNS%
\begin{equation}
L_{-2}\left\vert n-2,0,q\right\rangle =\frac{1}{2}\left\vert
n,0,q\right\rangle +M\left\vert n,0,q+1\right\rangle \label{9}%
\end{equation}
gives the second high-energy Ward identities%
\begin{equation}
\mathcal{T}^{(n,0,q)}=(-\frac{1}{2M})^{q}\mathcal{T}^{(n,0,0)}. \label{10}%
\end{equation}
Combining Eqs.(\ref{8}) and (\ref{10}) gives the master formula
\cite{CHLTY1,CHLTY2,CHLTY3}
\begin{equation}
\mathcal{T}^{(n,2m,q)}=(-\frac{1}{M})^{2m+q}(\frac{1}{2})^{m+q}%
(2m-1)!!\mathcal{T}^{(n,0,0)}, \label{11}%
\end{equation}
which shows that there is only one independent high-energy scattering
amplitudes at each fixed mass level.

We now consider the superstring case. We will first consider high-energy
scattering amplitudes of string states with polarizations on the scattering
plane. Those with polariations orthogonal to the scattering plane will be
discussed in section V. It can be argued that there are four types of
high-energy scattering amplitudes for states in the NS sector with even GSO
parity%
\begin{equation}
\left\vert n,2m,q\right\rangle \otimes\left\vert b_{-\frac{1}{2}}%
^{T}\right\rangle \equiv(\alpha_{-1}^{T})^{n-2m-2q}(\alpha_{-1}^{L}%
)^{2m}(\alpha_{-2}^{L})^{q}(b_{-\frac{1}{2}}^{T})\left\vert 0,k\right\rangle ,
\label{12}%
\end{equation}%
\begin{equation}
\left\vert n,2m+1,q\right\rangle \otimes\left\vert b_{-\frac{1}{2}}%
^{L}\right\rangle \equiv(\alpha_{-1}^{T})^{n-2m-2q-1}(\alpha_{-1}^{L}%
)^{2m+1}(\alpha_{-2}^{L})^{q}(b_{-\frac{1}{2}}^{L})\left\vert 0,k\right\rangle
, \label{13}%
\end{equation}%
\begin{equation}
\left\vert n,2m,q\right\rangle \otimes\left\vert b_{-\frac{3}{2}}%
^{L}\right\rangle \equiv(\alpha_{-1}^{T})^{n-2m-2q}(\alpha_{-1}^{L}%
)^{2m}(\alpha_{-2}^{L})^{q}(b_{-\frac{3}{2}}^{L})\left\vert 0,k\right\rangle ,
\label{14}%
\end{equation}%
\begin{equation}
\left\vert n,2m,q\right\rangle \otimes\left\vert b_{-\frac{1}{2}}^{T}%
b_{-\frac{1}{2}}^{L}b_{-\frac{3}{2}}^{L}\right\rangle \equiv(\alpha_{-1}%
^{T})^{n-2m-2q}(\alpha_{-1}^{L})^{2m}(\alpha_{-2}^{L})^{q}(b_{-\frac{1}{2}%
}^{T})(b_{-\frac{1}{2}}^{L})(b_{-\frac{3}{2}}^{L})\left\vert 0,k\right\rangle
. \label{15}%
\end{equation}
Note that the number of $\alpha_{-1}^{L}$ operator in Eq.(\ref{13}) is odd. In
the OCFQ spectrum of open superstring, the solutions of physical states
conditions include positive-norm propagating states and two types of zero-norm
states. In the NS sector, the latter are \cite{GSW}%
\begin{equation}
\text{Type I}:G_{-\frac{1}{2}}\left\vert x\right\rangle ,\text{ where
}G_{\frac{1}{2}}\left\vert x\right\rangle =G_{\frac{3}{2}}\left\vert
x\right\rangle =0,\text{ }L_{0}\left\vert x\right\rangle =0; \label{16}%
\end{equation}%
\begin{equation}
\text{Type II}:(G_{-\frac{3}{2}}+2G_{-\frac{1}{2}}L_{-1})\left\vert
\widetilde{x}\right\rangle ,\text{ where }G_{\frac{1}{2}}\left\vert
\widetilde{x}\right\rangle =G_{\frac{3}{2}}\left\vert \widetilde
{x}\right\rangle =0,\text{ }(L_{0}+1)\left\vert \widetilde{x}\right\rangle =0.
\label{17}%
\end{equation}
While Type I states have zero-norm at any space-time dimension, Type II states
have zero-norm \emph{only} at D=10. We will show that, for each fixed mass
level, all high-energy scattering amplitudes corresponding to states in
Eqs.(\ref{12})-(\ref{15}) are proportional to each other, and the
proportionality constants can be determined from the decoupling of two types
of zero-norm states, Eqs.(\ref{16}) and (\ref{17}) in the high-energy limit.
For simplicity, based on the result of Eq.(\ref{11}), one needs only calculate
the proportionality constants among the scattering amplitudes of the following
four lower mass level states%
\begin{equation}
\left\vert 2,0,0\right\rangle \otimes\left\vert b_{-\frac{1}{2}}%
^{T}\right\rangle \equiv(\alpha_{-1}^{T})^{2}(b_{-\frac{1}{2}}^{T})\left\vert
0,k\right\rangle , \label{18}%
\end{equation}%
\begin{equation}
\left\vert 2,1,0\right\rangle \otimes\left\vert b_{-\frac{1}{2}}%
^{L}\right\rangle \equiv(\alpha_{-1}^{T})(\alpha_{-1}^{L})(b_{-\frac{1}{2}%
}^{L})\left\vert 0,k\right\rangle , \label{19}%
\end{equation}%
\begin{equation}
\left\vert 1,0,0\right\rangle \otimes\left\vert b_{-\frac{3}{2}}%
^{L}\right\rangle \equiv(\alpha_{-1}^{T})(b_{-\frac{3}{2}}^{L})\left\vert
0,k\right\rangle , \label{20}%
\end{equation}%
\begin{equation}
\left\vert 0,0,0\right\rangle \otimes\left\vert b_{-\frac{1}{2}}^{T}%
b_{-\frac{1}{2}}^{L}b_{-\frac{3}{2}}^{L}\right\rangle \equiv(b_{-\frac{1}{2}%
}^{T})(b_{-\frac{1}{2}}^{L})(b_{-\frac{3}{2}}^{L})\left\vert 0,k\right\rangle
. \label{21}%
\end{equation}
Other proportionality constants for higher mass level can be obtained through
Eqs.(\ref{11}) and (\ref{18})-(\ref{21}). To calculate the ratio among the
high-energy scattering amplitudes corresponding to states in Eqs.(\ref{19})
and (\ref{20}), we use the decoupling of the Type I HZNS at mass level
$M^{2}=2$
\begin{equation}
G_{-\frac{1}{2}}(\alpha_{-1}^{L})\left\vert 0,k\right\rangle =[M(\alpha
_{-1}^{L})(b_{-\frac{1}{2}}^{L})+(b_{-\frac{3}{2}}^{L})]\left\vert
0,k\right\rangle . \label{22}%
\end{equation}
Eq.(\ref{22}) gives the ratio for states at mass level $M^{2}=4$
\begin{equation}
(\alpha_{-1}^{T})(b_{-\frac{3}{2}}^{L})\left\vert 0,k\right\rangle
:(\alpha_{-1}^{T})(\alpha_{-1}^{L})(b_{-\frac{1}{2}}^{L})\left\vert
0,k\right\rangle =M:-1. \label{23}%
\end{equation}
We have used an abbreviated notation for the scattering amplitudes on the
l.h.s. of Eq.(\ref{23}). The HZNS in Eq.(\ref{22}) is the high-energy limit of
the vector zero-norm state at mass level $M^{2}=2$%
\begin{equation}
G_{-\frac{1}{2}}\left\vert x\right\rangle =[k_{(\mu}\theta_{\nu)}\alpha
_{-1}^{\mu}b_{-\frac{1}{2}}^{\nu}+\theta\cdot b_{-\frac{3}{2}}]\left\vert
0,k\right\rangle , \label{24}%
\end{equation}
where%
\begin{equation}
\left\vert x\right\rangle =[\theta\cdot\alpha_{-1}+\frac{1}{2}k\cdot
b_{-\frac{1}{2}}\theta\cdot b_{-\frac{1}{2}}]\left\vert 0,k\right\rangle
,k\cdot\theta=0 \label{25}%
\end{equation}
In fact, in the high-energy limit, $\theta=$ $e^{L},$ so $\left\vert
x\right\rangle \rightarrow(\alpha_{-1}^{L})\left\vert 0,k\right\rangle $ and
Eq.(\ref{24}) reduces to Eq.(\ref{22}). To calculate the ratio among the
high-energy scattering amplitudes corresponding to states in Eqs.(\ref{18})
and (\ref{20}), we use the decoupling of the Type II HZNS at mass level
$M^{2}=4$
\begin{equation}
G_{-\frac{3}{2}}(\alpha_{-1}^{T})\left\vert 0,k\right\rangle =[M(\alpha
_{-1}^{T})(b_{-\frac{3}{2}}^{L})+(\alpha_{-1}^{T})^{2}(b_{-\frac{1}{2}}%
^{T})]\left\vert 0,k\right\rangle . \label{26}%
\end{equation}
Eq.(\ref{26}) gives the ratio%
\begin{equation}
(\alpha_{-1}^{T})(b_{-\frac{3}{2}}^{L})\left\vert 0,k\right\rangle
:(\alpha_{-1}^{T})^{2}(b_{-\frac{1}{2}}^{T})\left\vert 0,k\right\rangle =1:-M.
\label{27}%
\end{equation}
Finally, To calculate the ratio among the high-energy scattering amplitudes
corresponding to states in Eqs.(\ref{18}) and (\ref{21}), we use the
decoupling of the Type II HZNS at mass level $M^{2}=4$
\begin{equation}
G_{-\frac{3}{2}}(b_{-\frac{1}{2}}^{T})(b_{-\frac{1}{2}}^{L})\left\vert
0,k\right\rangle \equiv\lbrack M(b_{-\frac{1}{2}}^{T})(b_{-\frac{1}{2}}%
^{L})(b_{-\frac{3}{2}}^{L})+(\alpha_{-2}^{L})(b_{-\frac{1}{2}}^{T})]\left\vert
0,k\right\rangle . \label{28}%
\end{equation}
Eq.(\ref{28}) gives the ratio%
\begin{equation}
(b_{-\frac{1}{2}}^{T})(b_{-\frac{1}{2}}^{L})(b_{-\frac{3}{2}}^{L})\left\vert
0,k\right\rangle :(\alpha_{-2}^{L})(b_{-\frac{1}{2}}^{T})\left\vert
0,k\right\rangle =1:-M. \label{29}%
\end{equation}
On the other hand, Eq.(\ref{10}) gives%
\begin{equation}
(\alpha_{-2}^{L})(b_{-\frac{1}{2}}^{T})\left\vert 0,k\right\rangle
:(\alpha_{-1}^{T})^{2}(b_{-\frac{1}{2}}^{T})\left\vert 0,k\right\rangle
=1:-2M. \label{30}%
\end{equation}
We conclude that%
\begin{equation}
(b_{-\frac{1}{2}}^{T})(b_{-\frac{1}{2}}^{L})(b_{-\frac{3}{2}}^{L})\left\vert
0,k\right\rangle :(\alpha_{-1}^{T})^{2}(b_{-\frac{1}{2}}^{T})\left\vert
0,k\right\rangle =1:2M^{2}. \label{31}%
\end{equation}
Eqs.(\ref{23}),(\ref{27}) and (\ref{31}) give the proportionality constants
among high-energy scattering amplitudes corresponding to states in
Eqs.(\ref{18})-(\ref{21}). Finally, by using Eq.(\ref{11}), one can then
easily calculate the proportionality constants among high-energy scattering
amplitudes corresponding to states in Eqs.(\ref{12})-(\ref{15}).

\section{Virasoro Constraints}

In this section, we will use the method of Virasoro constrains to derive the
ratios between the physical states in the NS sector. In the superstring
theory, the physical state $\left\vert \phi\right\rangle $ in the NS sector
should satisfy the following conditions:
\begin{align}
\left(  L_{0}-\dfrac{1}{2}\right)  \left\vert \phi\right\rangle  &  =0,\\
L_{m}\left\vert \phi\right\rangle  &  =0\text{, }m=1,2,3,\cdots,\label{Lm}\\
G_{r}\left\vert \phi\right\rangle  &  =0\text{, }r=\dfrac{1}{2},\dfrac{3}%
{2},\dfrac{5}{2},\cdots, \label{Gr}%
\end{align}
where the $L_{m}$ and $G_{r}$ are super Virasoro operators in the NS sector,%
\begin{align}
L_{m}  &  =\frac{1}{2}\sum_{n}:\alpha_{m-n}\cdot\alpha_{n}:+\frac{1}{4}%
\sum_{r}\left(  2r-m\right)  :\psi_{m-r}\cdot\psi_{r}:,\\
G_{r}  &  =\sum_{n}\alpha_{n}\cdot\psi_{r-n}.
\end{align}
These super Virasoro operators satisfy the following superconformal algebra,%
\begin{align}
\left[  L_{m},L_{n}\right]   &  =\left(  m-n\right)  L_{m+n}+\dfrac{1}%
{8}D\left(  m^{3}-m\right)  \delta_{m+n},\nonumber\\
\left[  L_{m},G_{r}\right]   &  =\left(  \dfrac{1}{2}m-r\right)
G_{m+r},\nonumber\\
\left\{  G_{r},G_{s}\right\}   &  =2L_{r+s}+\dfrac{1}{2}D\left(  r^{2}%
-\dfrac{1}{4}\right)  \delta_{r+s}.
\end{align}
Using the above superconformal algebra, the Virasoro conditions (\ref{Lm}) and
(\ref{Gr}) reduce to the following simple form,%
\begin{align}
G_{1/2}\left\vert \phi\right\rangle  &  =0,\label{G1/2}\\
G_{3/2}\left\vert \phi\right\rangle  &  =0. \label{G3/2}%
\end{align}
In the following, we will use the reduced Virasoro conditions (\ref{G1/2}) and
(\ref{G3/2}) to determine the ratios between the physical states in the NS
sector in the high-energy limit.

To warm up, let us consider the mass level at $M^{2}=2$ first. The most
general state in the NS sector at this mass level can be written as%
\begin{equation}
\left\vert 2\right\rangle =\left\{
\begin{tabular}
[c]{|c|}\hline
$\mu$\\\hline
\end{tabular}
\psi_{-\frac{3}{2}}^{\mu}+%
\begin{tabular}
[c]{|c|}\hline
$\mu$\\\hline
\end{tabular}
\otimes%
\begin{tabular}
[c]{|c|}\hline
$\nu$\\\hline
\end{tabular}
\alpha_{-1}^{\mu}\psi_{-\frac{1}{2}}^{\nu}+%
\begin{tabular}
[c]{|c|}\hline
$\mu$\\\hline
$\nu$\\\hline
$\sigma$\\\hline
\end{tabular}
\psi_{-\frac{1}{2}}^{\mu}\psi_{-\frac{1}{2}}^{\nu}\psi_{-\frac{1}{2}}^{\sigma
}\right\}  \left\vert 0\right\rangle _{NS}, \label{|2>}%
\end{equation}
where we use the Young tableaux to represent the coefficients of different
tensors. The properties of symmetry and anti-symmetry can be easily and
clearly described in this representation.

We then apply the reduced Virasoro conditions (\ref{G1/2}) and (\ref{G3/2}) to
the state (\ref{|2>}). It is easy to obtain%
\begin{subequations}%
\begin{align}
G_{1/2}\left\vert 2\right\rangle  &  =\alpha_{-1}^{\mu}\left\{
\begin{tabular}
[c]{|c|}\hline
$\mu$\\\hline
\end{tabular}
+k^{\nu}%
\begin{tabular}
[c]{|c|}\hline
$\mu$\\\hline
\end{tabular}
\otimes%
\begin{tabular}
[c]{|c|}\hline
$\nu$\\\hline
\end{tabular}
\right\}  +\psi_{-\frac{1}{2}}^{\mu}\psi_{-\frac{1}{2}}^{\nu}\left\{
\begin{tabular}
[c]{|c|}\hline
$\mu$\\\hline
\end{tabular}
\otimes%
\begin{tabular}
[c]{|c|}\hline
$\nu$\\\hline
\end{tabular}
-%
\begin{tabular}
[c]{|c|}\hline
$\nu$\\\hline
\end{tabular}
\otimes%
\begin{tabular}
[c]{|c|}\hline
$\mu$\\\hline
\end{tabular}
+3k^{\sigma}%
\begin{tabular}
[c]{|c|}\hline
$\mu$\\\hline
$\nu$\\\hline
$\sigma$\\\hline
\end{tabular}
\right\}  ,\\
G_{3/2}\left\vert 2\right\rangle  &  =%
\begin{tabular}
[c]{|c|}\hline
$\mu$\\\hline
\end{tabular}
k^{\mu}+%
\begin{tabular}
[c]{|c|}\hline
$\mu$\\\hline
\end{tabular}
\otimes%
\begin{tabular}
[c]{|c|}\hline
$\nu$\\\hline
\end{tabular}
\eta^{\mu\nu},
\end{align}%
\end{subequations}%
which leads to the following equations%
\begin{subequations}%
,%
\begin{align}%
\begin{tabular}
[c]{|c|}\hline
$\mu$\\\hline
\end{tabular}
+k^{\nu}%
\begin{tabular}
[c]{|c|}\hline
$\mu$\\\hline
\end{tabular}
\otimes%
\begin{tabular}
[c]{|c|}\hline
$\nu$\\\hline
\end{tabular}
&  =0,\\%
\begin{tabular}
[c]{|c|}\hline
$\mu$\\\hline
\end{tabular}
\otimes%
\begin{tabular}
[c]{|c|}\hline
$\nu$\\\hline
\end{tabular}
-%
\begin{tabular}
[c]{|c|}\hline
$\nu$\\\hline
\end{tabular}
\otimes%
\begin{tabular}
[c]{|c|}\hline
$\mu$\\\hline
\end{tabular}
+3k^{\sigma}%
\begin{tabular}
[c]{|c|}\hline
$\mu$\\\hline
$\nu$\\\hline
$\sigma$\\\hline
\end{tabular}
&  =0,\\%
\begin{tabular}
[c]{|c|}\hline
$\mu$\\\hline
\end{tabular}
k^{\mu}+%
\begin{tabular}
[c]{|c|}\hline
$\mu$\\\hline
\end{tabular}
\otimes%
\begin{tabular}
[c]{|c|}\hline
$\nu$\\\hline
\end{tabular}
\eta^{\mu\nu}  &  =0.
\end{align}%
\end{subequations}%
To solve the above equation, we first take the high-energy limit by letting
$\mu\rightarrow\left(  L,T\right)  $ and%
\begin{equation}
k^{\mu}\rightarrow M\left(  e^{L}\right)  ^{\mu}\text{, }\eta^{\mu\nu
}\rightarrow\left(  e^{T}\right)  ^{\mu}\left(  e^{T}\right)  ^{\nu}.
\end{equation}
The above equations reduce to%
\begin{align}%
\begin{tabular}
[c]{|c|}\hline
$\mu$\\\hline
\end{tabular}
+M%
\begin{tabular}
[c]{|c|}\hline
$\mu$\\\hline
\end{tabular}
\otimes%
\begin{tabular}
[c]{|c|}\hline
$L$\\\hline
\end{tabular}
&  =0,\\%
\begin{tabular}
[c]{|c|}\hline
$\mu$\\\hline
\end{tabular}
\otimes%
\begin{tabular}
[c]{|c|}\hline
$\nu$\\\hline
\end{tabular}
-%
\begin{tabular}
[c]{|c|}\hline
$\nu$\\\hline
\end{tabular}
\otimes%
\begin{tabular}
[c]{|c|}\hline
$\mu$\\\hline
\end{tabular}
&  =0,\\
M%
\begin{tabular}
[c]{|c|}\hline
$L$\\\hline
\end{tabular}
+%
\begin{tabular}
[c]{|c|}\hline
$T$\\\hline
\end{tabular}
\otimes%
\begin{tabular}
[c]{|c|}\hline
$T$\\\hline
\end{tabular}
&  =0.
\end{align}
At this mass level, the terms with odd number of $T$'s will be sub-leading in
the high-energy limit and be ignored, the resulting equations contain only
terms will even number of $T$' as following,
\begin{align}%
\begin{tabular}
[c]{|c|}\hline
$L$\\\hline
\end{tabular}
+M%
\begin{tabular}
[c]{|c|}\hline
$L$\\\hline
\end{tabular}
\otimes%
\begin{tabular}
[c]{|c|}\hline
$L$\\\hline
\end{tabular}
&  =0,\\
M%
\begin{tabular}
[c]{|c|}\hline
$L$\\\hline
\end{tabular}
+%
\begin{tabular}
[c]{|c|}\hline
$T$\\\hline
\end{tabular}
\otimes%
\begin{tabular}
[c]{|c|}\hline
$T$\\\hline
\end{tabular}
&  =0.
\end{align}
The ratio of the coefficients then can be obtained as

\begin{center}%
\begin{equation}%
\begin{tabular}
[c]{|l|c|}\hline
$\varepsilon_{TT}$ & $M^{2}\left(  =2\right)  $\\\hline
$\varepsilon_{LL}$ & $1$\\\hline
$\varepsilon_{L}$ & $-M\left(  =-\sqrt{2}\right)  $\\\hline
\end{tabular}
\label{ratio}%
\end{equation}

\end{center}

In the following, we will consider the general mass level at $M^{2}=\left(
2n-1\right)  $. At this mass level, the most general state can be written as%
\begin{equation}
\left\vert n\right\rangle =\left\{  \sum_{m_{j},m_{r}}\overset{n}%
{\underset{j=1}{\otimes}}\frac{1}{j^{m_{j}}m_{j}!}%
\begin{tabular}
[c]{|c|c|c|}\hline
$\mu_{1}^{j}$ & $\cdots$ & $\mu_{m_{j}}^{j}$\\\hline
\end{tabular}
\alpha_{-j}^{\mu_{1}^{j}\cdots\mu_{m_{j}}^{j}}\overset{n-1/2}{\underset
{r=1/2}{\otimes}}\frac{1}{m_{r}!}%
\begin{tabular}
[c]{|c|c|c|}\hline
$\nu_{1}^{r}$ & $\cdots$ & $\nu_{m_{r}}^{r}$\\\hline
\end{tabular}
^{T}\psi_{-r}^{\nu_{1}^{r}\cdots\nu_{m_{r}}^{r}}\right\}  \left\vert
0,k\right\rangle , \label{|n>}%
\end{equation}
where%
\begin{equation}%
\begin{tabular}
[c]{|c|c|c|}\hline
$\nu_{1}^{r}$ & $\cdots$ & $\nu_{m_{r}}^{r}$\\\hline
\end{tabular}
^{T}=%
\begin{tabular}
[c]{|c|}\hline
$\nu_{1}^{r}$\\\hline
$\vdots$\\\hline
$\nu_{m_{r}}^{r}$\\\hline
\end{tabular}
,
\end{equation}
and we have defined the abbreviation%
\begin{equation}
\alpha_{-j}^{\mu_{1}^{j}\cdots\mu_{m_{j}}^{j}}\equiv\alpha_{-j}^{\mu_{1}^{j}%
}\cdots\alpha_{-j}^{\mu_{m_{j}}^{j}}\text{ and }\psi_{-r}^{\nu_{1}^{r}%
\cdots\nu_{m_{r}}^{r}}\equiv\psi_{-r}^{\nu_{1}^{r}}\cdots\psi_{-r}^{\nu
_{m_{r}}^{r}},
\end{equation}
with $m_{j}\left(  m_{r}\right)  $ is the number of the operator $\alpha
_{-j}^{\mu}\left(  \psi_{-r}^{\nu}\right)  $ for $j\in Z$ and $r\in Z+1/2$.
The summation runs over all possible $m_{j}\left(  m_{r}\right)  $ with the
constrain%
\begin{equation}
\sum_{j=1}^{n}jm_{j}+\sum_{r=1/2}^{n-1/2}rm_{r}=n-\frac{1}{2}\text{ with
}m_{j},m_{r}\geq0,
\end{equation}
so that the total mass square is $2\left(  n-1\right)  $.

Next, we will apply the reduced Virasoro conditions (\ref{G1/2}) and
(\ref{G3/2}) to the state (\ref{|n>}),%
\begin{subequations}%
\begin{align}
G_{1/2}\left\vert n\right\rangle  &  =\sum_{m_{j}}\left[  k^{\nu_{1}^{1/2}}%
\begin{tabular}
[c]{|c|c|c|}\hline
$\nu_{1}^{1/2}$ & $\cdots$ & $\nu_{m_{1/2}}^{1/2}$\\\hline
\end{tabular}
\ ^{T}\right.  \overset{k}{\underset{j=1}{\otimes}}%
\begin{tabular}
[c]{|c|c|c|}\hline
$\mu_{1}^{j}$ & $\cdots$ & $\mu_{m_{j}}^{j}$\\\hline
\end{tabular}
\ \overset{s}{\underset{r\neq1/2}{\otimes}}%
\begin{tabular}
[c]{|c|c|c|}\hline
$\nu_{1}^{r}$ & $\cdots$ & $\nu_{m_{r}}^{r}$\\\hline
\end{tabular}
\ ^{T}\nonumber\\
&  +\sum_{l\geq1}\sum_{i=1}^{m_{l}}l%
\begin{tabular}
[c]{|c|c|c|c|c|}\hline
$\mu_{1}^{l}$ & $\cdots$ & $\hat{\mu}_{i}^{l}$ & $\cdots$ & $\mu_{m_{l}}^{l}%
$\\\hline
\end{tabular}
\ \otimes%
\begin{tabular}
[c]{|c|c|c|c|}\hline
$\mu_{i}^{l}$ & $\nu_{1}^{l+1/2}$ & $\cdots$ & $\nu_{m_{l+1/2}}^{l+1/2}%
$\\\hline
\end{tabular}
\ ^{T}\nonumber\\
&  \otimes%
\begin{tabular}
[c]{|c|c|c|}\hline
$\nu_{2}^{1/2}$ & $\cdots$ & $\nu_{m_{1/2}}^{1/2}$\\\hline
\end{tabular}
\ ^{T}\overset{k}{\underset{j\neq l}{\otimes}}%
\begin{tabular}
[c]{|c|c|c|}\hline
$\mu_{1}^{j}$ & $\cdots$ & $\mu_{m_{j}}^{j}$\\\hline
\end{tabular}
\ \overset{s}{\underset{r\neq1/2,l+1/2}{\otimes}}%
\begin{tabular}
[c]{|c|c|c|}\hline
$\nu_{1}^{r}$ & $\cdots$ & $\nu_{m_{r}}^{r}$\\\hline
\end{tabular}
\ ^{T}\nonumber\\
&  +\sum_{i=2}^{m_{1/2}}%
\begin{tabular}
[c]{|c|c|c|c|}\hline
$\nu_{i}^{1/2}$ & $\mu_{1}^{1}$ & $\cdots$ & $\mu_{m_{1}}^{1}$\\\hline
\end{tabular}
\ \otimes\left(  -1\right)  ^{i+1}%
\begin{tabular}
[c]{|c|c|c|c|c|}\hline
$\nu_{2}^{1/2}$ & $\cdots$ & $\hat{\nu}_{i}^{1/2}$ & $\cdots$ & $\nu_{m_{1/2}%
}^{1/2}$\\\hline
\end{tabular}
\ ^{T}\nonumber\\
&  \overset{k}{\underset{j\neq l}{\otimes}}%
\begin{tabular}
[c]{|c|c|c|}\hline
$\mu_{1}^{j}$ & $\cdots$ & $\mu_{m_{j}}^{j}$\\\hline
\end{tabular}
\ \overset{s}{\underset{r\neq1/2}{\otimes}}%
\begin{tabular}
[c]{|c|c|c|}\hline
$\nu_{1}^{r}$ & $\cdots$ & $\nu_{m_{r}}^{r}$\\\hline
\end{tabular}
\ ^{T}\nonumber\\
&  +\sum_{l\geq2}\sum_{i=1}^{m_{l-1/2}}\left(  -1\right)  ^{i+1}%
\begin{tabular}
[c]{|c|c|c|c|}\hline
$\nu_{i}^{l-1/2}$ & $\mu_{1}^{l}$ & $\cdots$ & $\mu_{m_{l}}^{l}$\\\hline
\end{tabular}
\ \otimes%
\begin{tabular}
[c]{|c|c|c|c|c|}\hline
$\nu_{1}^{l-1/2}$ & $\cdots$ & $\hat{\nu}_{i}^{l-1/2}$ & $\cdots$ &
$\nu_{m_{l-1/2}}^{l-1/2}$\\\hline
\end{tabular}
\ ^{T}\nonumber\\
&  \otimes%
\begin{tabular}
[c]{|c|c|c|}\hline
$\nu_{2}^{1/2}$ & $\cdots$ & $\nu_{m_{1/2}}^{1/2}$\\\hline
\end{tabular}
\ ^{T}\left.  \overset{k}{\underset{j\neq l}{\otimes}}%
\begin{tabular}
[c]{|c|c|c|}\hline
$\mu_{1}^{j}$ & $\cdots$ & $\mu_{m_{j}}^{j}$\\\hline
\end{tabular}
\ \overset{s}{\underset{r\neq1/2,l-1/2}{\otimes}}%
\begin{tabular}
[c]{|c|c|c|}\hline
$\nu_{1}^{r}$ & $\cdots$ & $\nu_{m_{r}}^{r}$\\\hline
\end{tabular}
\ ^{T}\right] \nonumber\\
&  \cdot\frac{1}{\left(  m_{1/2}-1\right)  !}\psi_{-1/2}^{\nu_{2}^{1/2}%
\cdots\nu_{m_{1/2}}^{1/2}}\prod_{j=1}^{k}\frac{1}{j^{m_{j}}m_{j}!}\alpha
_{-j}^{\mu_{1}^{j}\cdots\mu_{m_{j}}^{j}}\prod_{r\neq1/2}^{s}\frac{1}{m_{r}%
!}\psi_{-r}^{\nu_{1}^{r}\cdots\nu_{m_{r}}^{r}},
\end{align}
and%
\begin{align}
G_{3/2}\left\vert n\right\rangle  &  =\sum_{m_{j}}\left[
\begin{tabular}
[c]{|c|c|c|}\hline
$\nu_{1}^{3/2}$ & $\cdots$ & $\nu_{m_{3/2}}^{3/2}$\\\hline
\end{tabular}
\ ^{T}k^{\nu_{1}^{3/2}}\right.  \overset{k}{\underset{j=1}{\otimes}}%
\begin{tabular}
[c]{|c|c|c|}\hline
$\mu_{1}^{j}$ & $\cdots$ & $\mu_{m_{j}}^{j}$\\\hline
\end{tabular}
\ \overset{s}{\underset{r\neq3/2}{\otimes}}%
\begin{tabular}
[c]{|c|c|c|}\hline
$\nu_{1}^{r}$ & $\cdots$ & $\nu_{m_{r}}^{r}$\\\hline
\end{tabular}
\ ^{T}\nonumber\\
&  +\eta^{\mu\nu}%
\begin{tabular}
[c]{|c|c|c|c|}\hline
$\mu$ & $\mu_{1}^{1}$ & $\cdots$ & $\mu_{m_{1}}^{1}$\\\hline
\end{tabular}
\ \otimes%
\begin{tabular}
[c]{|c|c|c|c|}\hline
$\nu$ & $\nu_{1}^{1/2}$ & $\cdots$ & $\nu_{m_{1/2}}^{1/2}$\\\hline
\end{tabular}
\ ^{T}\otimes%
\begin{tabular}
[c]{|c|c|c|}\hline
$\nu_{2}^{3/2}$ & $\cdots$ & $\nu_{m_{3/2}}^{3/2}$\\\hline
\end{tabular}
\ ^{T}\nonumber\\
&  \overset{k}{\underset{j\neq1}{\otimes}}%
\begin{tabular}
[c]{|c|c|c|}\hline
$\mu_{1}^{j}$ & $\cdots$ & $\mu_{m_{j}}^{j}$\\\hline
\end{tabular}
\ \overset{s}{\underset{r\neq1/2,3/2}{\otimes}}%
\begin{tabular}
[c]{|c|c|c|}\hline
$\nu_{1}^{r}$ & $\cdots$ & $\nu_{m_{r}}^{r}$\\\hline
\end{tabular}
\ ^{T}\nonumber\\
&  +\sum_{l\geq1}\sum_{i=1}^{m_{l}}l%
\begin{tabular}
[c]{|c|c|c|c|c|}\hline
$\mu_{1}^{l}$ & $\cdots$ & $\hat{\mu}_{i}^{l}$ & $\cdots$ & $\mu_{m_{l}}^{l}%
$\\\hline
\end{tabular}
\ \otimes%
\begin{tabular}
[c]{|c|c|c|c|}\hline
$\mu_{i}^{l}$ & $\nu_{1}^{l+3/2}$ & $\cdots$ & $\nu_{m_{l+3/2}}^{l+3/2}%
$\\\hline
\end{tabular}
\ ^{T}\nonumber\\
&  \otimes%
\begin{tabular}
[c]{|c|c|c|}\hline
$\nu_{2}^{3/2}$ & $\cdots$ & $\nu_{m_{3/2}}^{3/2}$\\\hline
\end{tabular}
\ ^{T}\overset{k}{\underset{j\neq l}{\otimes}}%
\begin{tabular}
[c]{|c|c|c|}\hline
$\mu_{1}^{j}$ & $\cdots$ & $\mu_{m_{j}}^{j}$\\\hline
\end{tabular}
\ \overset{s}{\underset{r\neq3/2,l+3/2}{\otimes}}%
\begin{tabular}
[c]{|c|c|c|}\hline
$\nu_{1}^{r}$ & $\cdots$ & $\nu_{m_{r}}^{r}$\\\hline
\end{tabular}
\ ^{T}\nonumber\\
&  +\sum_{i=2}^{m_{3/2}}3%
\begin{tabular}
[c]{|c|c|c|c|}\hline
$\nu_{i}^{3/2}$ & $\mu_{1}^{3}$ & $\cdots$ & $\mu_{m_{3}}^{3}$\\\hline
\end{tabular}
\ \otimes\left(  -1\right)  ^{i+1}%
\begin{tabular}
[c]{|c|c|c|c|c|}\hline
$\nu_{2}^{3/2}$ & $\cdots$ & $\hat{\nu}_{i}^{3/2}$ & $\cdots$ & $\nu_{m_{3/2}%
}^{3/2}$\\\hline
\end{tabular}
\ ^{T}\nonumber\\
&  \overset{k}{\underset{j\neq3}{\otimes}}%
\begin{tabular}
[c]{|c|c|c|}\hline
$\mu_{1}^{j}$ & $\cdots$ & $\mu_{m_{j}}^{j}$\\\hline
\end{tabular}
\ \overset{s}{\underset{r\neq3/2}{\otimes}}%
\begin{tabular}
[c]{|c|c|c|}\hline
$\nu_{1}^{r}$ & $\cdots$ & $\nu_{m_{r}}^{r}$\\\hline
\end{tabular}
\ ^{T}\nonumber\\
&  +\sum_{l\geq2,l\neq3}\sum_{i=1}^{m_{l-3/2}}%
\begin{tabular}
[c]{|c|c|c|c|}\hline
$\nu_{i}^{l-3/2}$ & $\mu_{1}^{l}$ & $\cdots$ & $\mu_{m_{l}}^{l}$\\\hline
\end{tabular}
\ \otimes%
\begin{tabular}
[c]{|c|c|c|c|c|}\hline
$\nu_{1}^{l-3/2}$ & $\cdots$ & $\hat{\nu}_{i}^{l-3/2}$ & $\cdots$ &
$\nu_{m_{l-3/2}}^{l-3/2}$\\\hline
\end{tabular}
\ ^{T}\nonumber\\
&  \otimes%
\begin{tabular}
[c]{|c|c|c|}\hline
$\nu_{2}^{3/2}$ & $\cdots$ & $\nu_{m_{3/2}}^{3/2}$\\\hline
\end{tabular}
\ ^{T}\left.  \overset{k}{\underset{j\neq l}{\otimes}}%
\begin{tabular}
[c]{|c|c|c|}\hline
$\mu_{1}^{j}$ & $\cdots$ & $\mu_{m_{j}}^{j}$\\\hline
\end{tabular}
\ \overset{s}{\underset{r\neq3/2,l-3/2}{\otimes}}%
\begin{tabular}
[c]{|c|c|c|}\hline
$\nu_{1}^{r}$ & $\cdots$ & $\nu_{m_{r}}^{r}$\\\hline
\end{tabular}
\ ^{T}\right] \nonumber\\
&  \cdot\frac{1}{\left(  m_{3/2}-1\right)  !}\psi_{-3/2}^{\nu_{2}^{3/2}%
\cdots\nu_{m_{3/2}}^{3/2}}\prod_{j=1}^{k}\frac{1}{j^{m_{j}}m_{j}!}\alpha
_{-j}^{\mu_{1}^{j}\cdots\mu_{m_{j}}^{j}}\prod_{r\neq3/2}^{s}\frac{1}{m_{r}%
!}\psi_{-r}^{\nu_{1}^{r}\cdots\nu_{m_{r}}^{r}},
\end{align}%
\end{subequations}%
where we have used the identities of the Young tableaux,%
\begin{align}%
\begin{tabular}
[c]{|c|c|c|}\hline
$1$ & $\cdots$ & $p$\\\hline
\end{tabular}
\  &  =\frac{1}{p}\left[  1+\sigma_{\left(  21\right)  }+\sigma_{\left(
321\right)  }+\cdots+\sigma_{\left(  p\cdots1\right)  }\right]
\begin{tabular}
[c]{|c|}\hline
$1$\\\hline
\end{tabular}
\ \otimes%
\begin{tabular}
[c]{|c|c|c|}\hline
$2$ & $\cdots$ & $p$\\\hline
\end{tabular}
\nonumber\\
&  =\frac{1}{p}\sum_{i=1}^{p}\sigma_{\left(  i1\right)  }%
\begin{tabular}
[c]{|c|}\hline
$1$\\\hline
\end{tabular}
\ \otimes%
\begin{tabular}
[c]{|c|c|c|}\hline
$2$ & $\cdots$ & $p$\\\hline
\end{tabular}
\ ,\\%
\begin{tabular}
[c]{|c|c|c|}\hline
$1$ & $\cdots$ & $p$\\\hline
\end{tabular}
\ ^{T}  &  =\frac{1}{p}\left[  1-\sigma_{\left(  21\right)  }+\sigma_{\left(
321\right)  }-\cdots+\left(  -1\right)  ^{p+1}\sigma_{\left(  p\cdots1\right)
}\right]
\begin{tabular}
[c]{|c|}\hline
$1$\\\hline
\end{tabular}
\ \otimes%
\begin{tabular}
[c]{|c|c|c|}\hline
$2$ & $\cdots$ & $p$\\\hline
\end{tabular}
\ ^{T}\nonumber\\
&  =\frac{1}{p}\sum_{i=1}^{p}\left(  -1\right)  ^{i+1}\sigma_{\left(
i\cdots1\right)  }%
\begin{tabular}
[c]{|c|}\hline
$1$\\\hline
\end{tabular}
\ \otimes%
\begin{tabular}
[c]{|c|c|c|}\hline
$2$ & $\cdots$ & $p$\\\hline
\end{tabular}
\ ^{T}.
\end{align}
We then obtain the constraint equations%
\begin{subequations}%
\begin{align}
0  &  =k^{\nu_{1}^{1/2}}%
\begin{tabular}
[c]{|c|c|c|}\hline
$\nu_{1}^{1/2}$ & $\cdots$ & $\nu_{m_{1/2}}^{1/2}$\\\hline
\end{tabular}
\ ^{T}\overset{k}{\underset{j=1}{\otimes}}%
\begin{tabular}
[c]{|c|c|c|}\hline
$\mu_{1}^{j}$ & $\cdots$ & $\mu_{m_{j}}^{j}$\\\hline
\end{tabular}
\ \overset{s}{\underset{r\neq1/2}{\otimes}}%
\begin{tabular}
[c]{|c|c|c|}\hline
$\nu_{1}^{r}$ & $\cdots$ & $\nu_{m_{r}}^{r}$\\\hline
\end{tabular}
\ ^{T}\nonumber\\
&  +\sum_{l\geq1}\sum_{i=1}^{m_{l}}l%
\begin{tabular}
[c]{|c|c|c|c|c|}\hline
$\mu_{1}^{l}$ & $\cdots$ & $\hat{\mu}_{i}^{l}$ & $\cdots$ & $\mu_{m_{l}}^{l}%
$\\\hline
\end{tabular}
\ \otimes%
\begin{tabular}
[c]{|c|c|c|c|}\hline
$\mu_{i}^{l}$ & $\nu_{1}^{l+1/2}$ & $\cdots$ & $\nu_{m_{l+1/2}}^{l+1/2}%
$\\\hline
\end{tabular}
\ ^{T}\nonumber\\
&  \otimes%
\begin{tabular}
[c]{|c|c|c|}\hline
$\nu_{2}^{1/2}$ & $\cdots$ & $\nu_{m_{1/2}}^{1/2}$\\\hline
\end{tabular}
\ ^{T}\overset{k}{\underset{j\neq l}{\otimes}}%
\begin{tabular}
[c]{|c|c|c|}\hline
$\mu_{1}^{j}$ & $\cdots$ & $\mu_{m_{j}}^{j}$\\\hline
\end{tabular}
\ \overset{s}{\underset{r\neq1/2,l+1/2}{\otimes}}%
\begin{tabular}
[c]{|c|c|c|}\hline
$\nu_{1}^{r}$ & $\cdots$ & $\nu_{m_{r}}^{r}$\\\hline
\end{tabular}
\ ^{T}\nonumber\\
&  +\sum_{i=2}^{m_{1/2}}%
\begin{tabular}
[c]{|c|c|c|c|}\hline
$\nu_{i}^{1/2}$ & $\mu_{1}^{1}$ & $\cdots$ & $\mu_{m_{1}}^{1}$\\\hline
\end{tabular}
\ \otimes\left(  -1\right)  ^{i+1}%
\begin{tabular}
[c]{|c|c|c|c|c|}\hline
$\nu_{2}^{1/2}$ & $\cdots$ & $\hat{\nu}_{i}^{1/2}$ & $\cdots$ & $\nu_{m_{1/2}%
}^{1/2}$\\\hline
\end{tabular}
\ ^{T}\nonumber\\
&  \overset{k}{\underset{j\neq l}{\otimes}}%
\begin{tabular}
[c]{|c|c|c|}\hline
$\mu_{1}^{j}$ & $\cdots$ & $\mu_{m_{j}}^{j}$\\\hline
\end{tabular}
\ \overset{s}{\underset{r\neq1/2}{\otimes}}%
\begin{tabular}
[c]{|c|c|c|}\hline
$\nu_{1}^{r}$ & $\cdots$ & $\nu_{m_{r}}^{r}$\\\hline
\end{tabular}
\ ^{T}\nonumber\\
&  +\sum_{l\geq2}\sum_{i=1}^{m_{l-1/2}}\left(  -1\right)  ^{i+1}%
\begin{tabular}
[c]{|c|c|c|c|}\hline
$\nu_{i}^{l-1/2}$ & $\mu_{1}^{l}$ & $\cdots$ & $\mu_{m_{l}}^{l}$\\\hline
\end{tabular}
\ \otimes%
\begin{tabular}
[c]{|c|c|c|c|c|}\hline
$\nu_{1}^{l-1/2}$ & $\cdots$ & $\hat{\nu}_{i}^{l-1/2}$ & $\cdots$ &
$\nu_{m_{l-1/2}}^{l-1/2}$\\\hline
\end{tabular}
\ ^{T}\nonumber\\
&  \otimes%
\begin{tabular}
[c]{|c|c|c|}\hline
$\nu_{2}^{1/2}$ & $\cdots$ & $\nu_{m_{1/2}}^{1/2}$\\\hline
\end{tabular}
\ ^{T}\overset{k}{\underset{j\neq l}{\otimes}}%
\begin{tabular}
[c]{|c|c|c|}\hline
$\mu_{1}^{j}$ & $\cdots$ & $\mu_{m_{j}}^{j}$\\\hline
\end{tabular}
\ \overset{s}{\underset{r\neq1/2,l-1/2}{\otimes}}%
\begin{tabular}
[c]{|c|c|c|}\hline
$\nu_{1}^{r}$ & $\cdots$ & $\nu_{m_{r}}^{r}$\\\hline
\end{tabular}
\ ^{T},
\end{align}%
\begin{align}
0  &  =%
\begin{tabular}
[c]{|c|c|c|}\hline
$\nu_{1}^{3/2}$ & $\cdots$ & $\nu_{m_{3/2}}^{3/2}$\\\hline
\end{tabular}
\ ^{T}k^{\nu_{1}^{3/2}}\overset{k}{\underset{j=1}{\otimes}}%
\begin{tabular}
[c]{|c|c|c|}\hline
$\mu_{1}^{j}$ & $\cdots$ & $\mu_{m_{j}}^{j}$\\\hline
\end{tabular}
\ \overset{s}{\underset{r\neq3/2}{\otimes}}%
\begin{tabular}
[c]{|c|c|c|}\hline
$\nu_{1}^{r}$ & $\cdots$ & $\nu_{m_{r}}^{r}$\\\hline
\end{tabular}
\ ^{T}\nonumber\\
&  +\eta^{\mu\nu}%
\begin{tabular}
[c]{|c|c|c|c|}\hline
$\mu$ & $\mu_{1}^{1}$ & $\cdots$ & $\mu_{m_{1}}^{1}$\\\hline
\end{tabular}
\ \otimes%
\begin{tabular}
[c]{|c|c|c|c|}\hline
$\nu$ & $\nu_{1}^{1/2}$ & $\cdots$ & $\nu_{m_{1/2}}^{1/2}$\\\hline
\end{tabular}
\ ^{T}\otimes%
\begin{tabular}
[c]{|c|c|c|}\hline
$\nu_{2}^{3/2}$ & $\cdots$ & $\nu_{m_{3/2}}^{3/2}$\\\hline
\end{tabular}
\ ^{T}\nonumber\\
&  \overset{k}{\underset{j\neq1}{\otimes}}%
\begin{tabular}
[c]{|c|c|c|}\hline
$\mu_{1}^{j}$ & $\cdots$ & $\mu_{m_{j}}^{j}$\\\hline
\end{tabular}
\ \overset{s}{\underset{r\neq1/2,3/2}{\otimes}}%
\begin{tabular}
[c]{|c|c|c|}\hline
$\nu_{1}^{r}$ & $\cdots$ & $\nu_{m_{r}}^{r}$\\\hline
\end{tabular}
\ ^{T}\nonumber\\
&  +\sum_{l\geq1}\sum_{i=1}^{m_{l}}l%
\begin{tabular}
[c]{|c|c|c|c|c|}\hline
$\mu_{1}^{l}$ & $\cdots$ & $\hat{\mu}_{i}^{l}$ & $\cdots$ & $\mu_{m_{l}}^{l}%
$\\\hline
\end{tabular}
\ \otimes%
\begin{tabular}
[c]{|c|c|c|c|}\hline
$\mu_{i}^{l}$ & $\nu_{1}^{l+3/2}$ & $\cdots$ & $\nu_{m_{l+3/2}}^{l+3/2}%
$\\\hline
\end{tabular}
\ ^{T}\nonumber\\
&  \otimes%
\begin{tabular}
[c]{|c|c|c|}\hline
$\nu_{2}^{3/2}$ & $\cdots$ & $\nu_{m_{3/2}}^{3/2}$\\\hline
\end{tabular}
\ ^{T}\overset{k}{\underset{j\neq l}{\otimes}}%
\begin{tabular}
[c]{|c|c|c|}\hline
$\mu_{1}^{j}$ & $\cdots$ & $\mu_{m_{j}}^{j}$\\\hline
\end{tabular}
\ \overset{s}{\underset{r\neq3/2,l+3/2}{\otimes}}%
\begin{tabular}
[c]{|c|c|c|}\hline
$\nu_{1}^{r}$ & $\cdots$ & $\nu_{m_{r}}^{r}$\\\hline
\end{tabular}
\ ^{T}\nonumber\\
&  +\sum_{i=2}^{m_{3/2}}3%
\begin{tabular}
[c]{|c|c|c|c|}\hline
$\nu_{i}^{3/2}$ & $\mu_{1}^{3}$ & $\cdots$ & $\mu_{m_{3}}^{3}$\\\hline
\end{tabular}
\ \otimes\left(  -1\right)  ^{i+1}%
\begin{tabular}
[c]{|c|c|c|c|c|}\hline
$\nu_{2}^{3/2}$ & $\cdots$ & $\hat{\nu}_{i}^{3/2}$ & $\cdots$ & $\nu_{m_{3/2}%
}^{3/2}$\\\hline
\end{tabular}
\ ^{T}\nonumber\\
&  \overset{k}{\underset{j\neq3}{\otimes}}%
\begin{tabular}
[c]{|c|c|c|}\hline
$\mu_{1}^{j}$ & $\cdots$ & $\mu_{m_{j}}^{j}$\\\hline
\end{tabular}
\ \overset{s}{\underset{r\neq3/2}{\otimes}}%
\begin{tabular}
[c]{|c|c|c|}\hline
$\nu_{1}^{r}$ & $\cdots$ & $\nu_{m_{r}}^{r}$\\\hline
\end{tabular}
\ ^{T}\nonumber\\
&  +\sum_{l\geq2,l\neq3}\sum_{i=1}^{m_{l-3/2}}%
\begin{tabular}
[c]{|c|c|c|c|}\hline
$\nu_{i}^{l-3/2}$ & $\mu_{1}^{l}$ & $\cdots$ & $\mu_{m_{l}}^{l}$\\\hline
\end{tabular}
\ \otimes\left(  -1\right)  ^{i+1}%
\begin{tabular}
[c]{|c|c|c|c|c|}\hline
$\nu_{1}^{l-3/2}$ & $\cdots$ & $\hat{\nu}_{i}^{l-3/2}$ & $\cdots$ &
$\nu_{m_{l-3/2}}^{l-3/2}$\\\hline
\end{tabular}
\ ^{T}\nonumber\\
&  \otimes%
\begin{tabular}
[c]{|c|c|c|}\hline
$\nu_{2}^{3/2}$ & $\cdots$ & $\nu_{m_{3/2}}^{3/2}$\\\hline
\end{tabular}
\ ^{T}\overset{k}{\underset{j\neq l}{\otimes}}%
\begin{tabular}
[c]{|c|c|c|}\hline
$\mu_{1}^{j}$ & $\cdots$ & $\mu_{m_{j}}^{j}$\\\hline
\end{tabular}
\ \overset{s}{\underset{r\neq3/2,l-3/2}{\otimes}}%
\begin{tabular}
[c]{|c|c|c|}\hline
$\nu_{1}^{r}$ & $\cdots$ & $\nu_{m_{r}}^{r}$\\\hline
\end{tabular}
\ ^{T}.
\end{align}%
\end{subequations}%
Taking the high-energy limit in the above equations by letting $\left(
\mu_{i},\nu_{i}\right)  \rightarrow\left(  L,T\right)  $, and%
\begin{equation}
k^{\mu_{i}}\rightarrow M\left(  e^{L}\right)  ^{\mu_{i}}\text{, }\eta^{\mu
_{1}\mu_{2}}\rightarrow\left(  e^{T}\right)  ^{\mu_{1}}\left(  e^{T}\right)
^{\mu_{2}},
\end{equation}
and using the following lemma,

\noindent\textbf{Lemma:}%
\begin{equation}%
\begin{tabular}
[c]{|l|l|l|}\hline
$T$ & $\cdots$ & $T$\\\hline
\end{tabular}
\underset{l_{1}}{\underbrace{%
\begin{tabular}
[c]{|l|l|l|}\hline
$L$ & $\cdots$ & $L$\\\hline
\end{tabular}
}}\otimes\underset{m_{2}-l_{2}}{\underbrace{%
\begin{tabular}
[c]{|l|l|l|}\hline
$T$ & $\cdots$ & $T$\\\hline
\end{tabular}
}}%
\begin{tabular}
[c]{|l|l|l|}\hline
$L$ & $\cdots$ & $L$\\\hline
\end{tabular}
\otimes\underset{m_{1/2}}{\underbrace{%
\begin{tabular}
[c]{|c|c|}\hline
$\nu_{1}^{1/2}$ & $\nu_{2}^{1/2}$\\\hline
\end{tabular}
}}\otimes%
\begin{tabular}
[c]{|c|}\hline
$\nu_{2}^{3/2}$\\\hline
\end{tabular}
\otimes%
\begin{tabular}
[c]{|lll|}\hline
& $\cdots$ & \\\hline
\end{tabular}
\equiv0, \label{lamma}%
\end{equation}
\textit{except for (i) }$m_{j\geq3}=m_{r\geq3/2}=0$\textit{, }$l_{2}=m_{2}%
$\textit{, }$l_{3/2}=m_{3/2}=1$\textit{ and (ii) }$l_{1}+l_{1/2}=2k$\textit{.}

\noindent we solve the equations in the appendix, the ratios between the
physical states in the NS sector in the high-energy limit are given as%
\begin{align}
&  \underset{n-2m_{2}-2k}{\underbrace{%
\begin{tabular}
[c]{|l|l|l|}\hline
$T$ & $\cdots$ & $T$\\\hline
\end{tabular}
\ }}\underset{2k}{\underbrace{%
\begin{tabular}
[c]{|l|l|l|}\hline
$L$ & $\cdots$ & $L$\\\hline
\end{tabular}
\ }}\otimes\underset{m_{2}}{\underbrace{%
\begin{tabular}
[c]{|l|l|l|}\hline
$L$ & $\cdots$ & $L$\\\hline
\end{tabular}
\ }}\otimes%
\begin{tabular}
[c]{|c|}\hline
$0$\\\hline
\end{tabular}
\otimes%
\begin{tabular}
[c]{|c|}\hline
$L$\\\hline
\end{tabular}
\nonumber\\
&  =\left(  -\frac{1}{2M}\right)  ^{m_{2}}\left(  -\frac{1}{2M}\right)
^{k}\frac{\left(  2k-1\right)  !!}{\left(  -M\right)  ^{k}}\underset
{n}{\underbrace{%
\begin{tabular}
[c]{|l|l|l|}\hline
$T$ & $\cdots$ & $T$\\\hline
\end{tabular}
\ }}\otimes%
\begin{tabular}
[c]{|c|}\hline
$0$\\\hline
\end{tabular}
\otimes%
\begin{tabular}
[c]{|c|}\hline
$0$\\\hline
\end{tabular}
\otimes%
\begin{tabular}
[c]{|c|}\hline
$L$\\\hline
\end{tabular}
,
\end{align}%
\begin{align}
&  \underset{n-2m_{2}-2k}{\underbrace{%
\begin{tabular}
[c]{|l|l|l|}\hline
$T$ & $\cdots$ & $T$\\\hline
\end{tabular}
\ \ }}\underset{2k+1}{\underbrace{%
\begin{tabular}
[c]{|l|l|l|}\hline
$L$ & $\cdots$ & $L$\\\hline
\end{tabular}
\ \ }}\otimes\underset{m_{2}}{\underbrace{%
\begin{tabular}
[c]{|l|l|l|}\hline
$L$ & $\cdots$ & $L$\\\hline
\end{tabular}
\ \ }}\otimes%
\begin{tabular}
[c]{|c|}\hline
$L$\\\hline
\end{tabular}
\otimes%
\begin{tabular}
[c]{|c|}\hline
$0$\\\hline
\end{tabular}
\nonumber\\
&  =\left(  -\frac{1}{2M}\right)  ^{m_{2}}\left(  -\frac{1}{2M}\right)
^{k}\frac{\left(  2k+1\right)  !!}{\left(  -M\right)  ^{k+1}}\underset
{n}{\underbrace{%
\begin{tabular}
[c]{|l|l|l|}\hline
$T$ & $\cdots$ & $T$\\\hline
\end{tabular}
\ \ }}\otimes%
\begin{tabular}
[c]{|c|}\hline
$0$\\\hline
\end{tabular}
\ \ \otimes%
\begin{tabular}
[c]{|c|}\hline
$0$\\\hline
\end{tabular}
\ \ \otimes%
\begin{tabular}
[c]{|c|}\hline
$L$\\\hline
\end{tabular}
\ \ ,
\end{align}%
\begin{align}
&  \underset{n-2m_{2}-2k+1}{\underbrace{%
\begin{tabular}
[c]{|l|l|l|}\hline
$T$ & $\cdots$ & $T$\\\hline
\end{tabular}
\ \ }}\underset{2k}{\underbrace{%
\begin{tabular}
[c]{|l|l|l|}\hline
$L$ & $\cdots$ & $L$\\\hline
\end{tabular}
\ \ }}\otimes\underset{m_{2}}{\underbrace{%
\begin{tabular}
[c]{|l|l|l|}\hline
$L$ & $\cdots$ & $L$\\\hline
\end{tabular}
\ \ }}\otimes%
\begin{tabular}
[c]{|c|}\hline
$T$\\\hline
\end{tabular}
\ \ \otimes%
\begin{tabular}
[c]{|c|}\hline
$0$\\\hline
\end{tabular}
\nonumber\\
&  =\left(  -\frac{1}{2M}\right)  ^{m_{2}}\left(  -\frac{1}{2M}\right)
^{k}\frac{\left(  2k-1\right)  !!}{\left(  -M\right)  ^{k-1}}\underset
{n}{\underbrace{%
\begin{tabular}
[c]{|l|l|l|}\hline
$T$ & $\cdots$ & $T$\\\hline
\end{tabular}
\ \ }}\otimes%
\begin{tabular}
[c]{|c|}\hline
$0$\\\hline
\end{tabular}
\ \ \otimes%
\begin{tabular}
[c]{|c|}\hline
$0$\\\hline
\end{tabular}
\ \ \otimes%
\begin{tabular}
[c]{|c|}\hline
$L$\\\hline
\end{tabular}
\ \ ,
\end{align}%
\begin{align}
&  \underset{n-2m_{2}-2k+1}{\underbrace{%
\begin{tabular}
[c]{|l|l|l|}\hline
$T$ & $\cdots$ & $T$\\\hline
\end{tabular}
\ \ }}\underset{2k}{\underbrace{%
\begin{tabular}
[c]{|l|l|l|}\hline
$L$ & $\cdots$ & $L$\\\hline
\end{tabular}
\ \ }}\otimes\underset{m_{2}-1}{\underbrace{%
\begin{tabular}
[c]{|l|l|l|}\hline
$L$ & $\cdots$ & $L$\\\hline
\end{tabular}
\ \ }}\otimes%
\begin{tabular}
[c]{|c|c|}\hline
$T$ & $L$\\\hline
\end{tabular}
\ \ ^{T}\otimes%
\begin{tabular}
[c]{|c|}\hline
$L$\\\hline
\end{tabular}
\nonumber\\
&  =\left(  -\frac{1}{2M}\right)  ^{m_{2}}\left(  -\frac{1}{2M}\right)
^{k}\frac{\left(  2k-1\right)  !!}{\left(  -M\right)  ^{k}}\underset
{n}{\underbrace{%
\begin{tabular}
[c]{|l|l|l|}\hline
$T$ & $\cdots$ & $T$\\\hline
\end{tabular}
\ \ }}\otimes%
\begin{tabular}
[c]{|c|}\hline
$0$\\\hline
\end{tabular}
\ \ \otimes%
\begin{tabular}
[c]{|c|}\hline
$0$\\\hline
\end{tabular}
\ \ \otimes%
\begin{tabular}
[c]{|c|}\hline
$L$\\\hline
\end{tabular}
\ \ ,
\end{align}
which are exactly consistent with the results obtained by using the decoupling
of HZNS in section II and the saddle-point calculation in the following section.

\section{Saddle-Point Approximation}

In this section, we shall calculate the high-energy limits of various
scattering amplitudes based on saddle-point approximation. Since the
decoupling of zero-norm states holds true for arbitrary physical processes, in
order to check the ratios among scattering amplitudes at the same mass level,
it is helpful to choose low-lying states to simplify calculations. For
instance, in the case of 4-point amplitudes, we fix the first vertex to be a
$M^{2}=0$ photon with polarization vector $\epsilon^{\mu}$ (in the $-1$ ghost
picture, and $\phi$ is the bosonized ghost operator),
\begin{equation}
V_{1}\equiv\epsilon^{\mu}\psi_{\mu}e^{-\phi}e^{ik_{1}X_{1}},\hspace
{1cm}\epsilon\cdot k_{1}=k_{1}^{2}=0; \label{63}%
\end{equation}
and the third and fourth vertices to be $M^{2}=-1$ tachyon (in the $0$ ghost
picture),
\begin{equation}
V_{3,4}\equiv k_{3,4}^{\mu}\psi_{\mu}e^{ik_{3,4}X_{3,4}},\hspace{1cm}%
k_{3,4}^{2}=-1. \label{64}%
\end{equation}
We shall vary the second vertex at the same level and compare the scattering
amplitudes to obtain the proportional constants.

\subsection{$M^{2} = 2$}

The second vertex operators at mass level $M^{2}=2$, are given by (in the $-1$
ghost picture),
\begin{align}
(\alpha_{-1}^{T})(b_{-\frac{1}{2}}^{T})\left\vert 0,k\right\rangle  &
\Rightarrow\psi^{T}\partial X^{T}e^{-\phi}e^{ikX},\label{65}\\
(\alpha_{-1}^{L})(b_{-\frac{1}{2}}^{L})\left\vert 0,k\right\rangle  &
\Rightarrow\psi^{L}\partial X^{L}e^{-\phi}e^{ikX},\label{66}\\
(b_{-\frac{3}{2}}^{L})\left\vert 0,k\right\rangle  &  \Rightarrow\partial
\psi^{L}e^{-\phi}e^{ikX}. \label{67}%
\end{align}
Here we have used the polarization basis to specify the particle spins,
e.g.,$\psi^{T}\equiv e_{\mu}^{T}\cdot\psi^{\mu}$.

To illustrate the procedure, we take the first state, Eq.(\ref{65}), as an
example to calculate the scattering amplitude among one massive tensor
$(M^{2}=2)$ with one photon $(V_{1})$ and two tachyons $(V_{3},V_{4})$. As in
the case of open bosonic string theory, we list the contributions of $s-t$
channel only. The 4-point function is given by
\begin{equation}
\int_{0}^{1}dx_{2}\langle(\psi_{1}^{T_{1}}e^{-\phi_{1}}e^{ik_{1}X_{1}}%
)(\psi_{2}^{T_{2}}\partial X_{2}^{T_{2}}e^{-\phi_{2}}e^{ik_{2}X_{2}%
})(k_{3\lambda}\psi_{3}^{\lambda}e^{ik_{3}X_{3}})(k_{4\sigma}\psi_{4}^{\sigma
}e^{ik_{4}X_{4}})\rangle, \label{68}%
\end{equation}
where we have suppressed the $SL(2,R)$ gauge-fixed world-sheet coordinates
$x_{1}=0,x_{3}=1,x_{4}=\infty$. Notice that in both the first and second
vertices, it is possible to allow fermion operators $\psi^{\mu}$ to have
polarization in transverse direction $T_{i}$ out of the scattering plane. As
we shall see in next section that this leads to a new feature of
supersymmetric stringy amplitudes in the high-energy limit. At this moment, we
only choose the polarization vector to be in the $P,L,T$ directions for a
comparison with results obtained by the previous two methods.

A direct application of Wick contraction among fermions $\psi$, ghosts $\phi$,
and bosons $X$ leads to the following result
\begin{equation}
\int_{0}^{1}dx\left[  \frac{(3,4)(e^{T_{1}}\cdot e^{T_{2}})}{x}-(e^{T_{1}%
}\cdot k_{3})(e^{T_{2}}\cdot k_{4})+\frac{(e^{T_{2}}\cdot k_{3})(e^{T_{1}%
}\cdot k_{4})}{1-x}\right]  \frac{1}{x}\left[  \frac{e^{T_{2}}\cdot k_{3}%
}{1-x}\right]  x^{(1,2)}(1-x)^{(2,3)}, \label{69}%
\end{equation}
where we have used the short-hand notation, $(3,4)\equiv k_{3}\cdot k_{4}$.
Based on the kinematic variables and the master formula for saddle-point
approximation,
\begin{align}
&  \int dx\hspace{0.3cm}u(x)\exp^{-Kf(x)}\nonumber\\
&  =u_{0}e^{-Kf_{0}}\sqrt{\frac{2\pi}{Kf_{0}^{\prime\prime}}}\left\{
1+\left[  \frac{u_{0}^{\prime\prime}}{2u_{0}f_{0}^{\prime\prime}}-\frac
{u_{0}^{\prime}f^{(3)}}{2u_{0}(f_{0}^{\prime\prime})^{2}}-\frac{f_{0}^{(4)}%
}{8(f_{0}^{\prime\prime})^{2}}+\frac{5[f^{(3)}]^{2}}{24(f_{0}^{\prime\prime
})^{3}}\right]  \frac{1}{K}+O(\frac{1}{{K}^{2}})\right\}  , \label{70}%
\end{align}
where $u_{0},f_{0},u_{0}^{\prime},f_{0}^{\prime\prime},$ etc, stand for the
values of functions and their derivatives evaluated at the saddle point
$f^{\prime}(x_{0})=0$. In order to apply this master formula to calculate
stringy amplitudes, we need the following substitutions $\left(
\alpha^{\prime}=1/2\right)  $
\begin{align}
K  &  \equiv2E^{2},\label{71}\\
f(x)  &  \equiv\ln(x)-\tau\ln(1-x),\label{72}\\
\tau &  \equiv-\frac{(2,3)}{(1,2)}\rightarrow\sin^{2}\frac{\theta}{2},
\label{73}%
\end{align}
where $\theta$ is the scattering angle in center of momentum frame and the
saddle point for the integration of moduli is $x_{0}=\frac{1}{1-\tau}$. In the
first scattering amplitude corresponding to Eq.(\ref{65}), we can identify the
$u(x)$ function as
\begin{equation}
u_{I}(x)\equiv\left[  \frac{(3,4)(e^{T_{1}}\cdot e^{T_{2}})}{x}-(e^{T_{1}%
}\cdot k_{3})(e^{T_{2}}\cdot k_{4})+\frac{(e^{T_{2}}\cdot k_{3})(e^{T_{1}%
}\cdot k_{4})}{1-x}\right]  \frac{1}{x}\left[  \frac{e^{T_{2}}\cdot k_{3}%
}{1-x}\right]  . \label{74}%
\end{equation}
Equipped with this, we obtain the high-energy limit of the first amplitude,
\begin{align}
&  2E^{2}(1-\tau)(e^{T}\cdot k_{3})x_{0}^{(1,2)-1}(1-x_{0})^{(2,3)-1}%
\sqrt{\frac{\pi\tau}{E^{2}(1-\tau)^{3}}}\nonumber\\
&  =4\sqrt{\pi}E^{2}(1-\tau)^{2}x_{0}^{(1,2)}(1-x_{0})^{(2,3)}. \label{75}%
\end{align}
Next, we replace the second vertex operator in Eq.(\ref{68}) by Eq.(\ref{66}),
and the 4-point function is given by
\begin{equation}
\int_{0}^{1}dx\frac{1}{M^{2}}\left[  (e^{T}\cdot k_{3})(2,4)-\frac{(e^{T}\cdot
k_{4})(2,3)}{1-x}\right]  \frac{1}{x}\left[  \frac{(1,2)}{x}-\frac{(2,3)}%
{1-x}\right]  x^{(1,2)}(1-x)^{(2,3)}. \label{76}%
\end{equation}
Here we can identify the $u(x)$ function for saddle-point master formula,
Eq.(\ref{70})
\begin{equation}
u_{II}(x)\equiv\frac{(e^{T}\cdot k_{3})(1,2)}{M^{2}x}\left[  (2,4)+\frac
{(2,3)}{1-x}\right]  f^{\prime}(x). \label{77}%
\end{equation}
One can check that $u_{II}(x_{0})=u_{II}^{\prime}(x_{0})=0$, and
\begin{equation}
u_{II}^{\prime\prime}(x_{0})=\frac{2(1,2)(2,3)(e^{T}\cdot k_{3})}%
{M^{2}x(1-x)^{2}}f^{\prime\prime}(x_{0}). \label{78}%
\end{equation}
Thus, the amplitude associated with the massive state, Eq.(\ref{66}), is given
by
\begin{align}
&  -\frac{2}{M^{2}}E^{2}\tau(e^{T}\cdot k_{3})x_{0}^{(1,2)-1}(1-x_{0}%
)^{(2,3)-2}\sqrt{\frac{\pi\tau}{E^{2}(1-\tau)^{3}}}\nonumber\\
&  =\frac{4}{M^{2}}\sqrt{\pi}E^{2}(1-\tau)^{2}x_{0}^{(1,2)}(1-x_{0})^{(2,3)}.
\label{79}%
\end{align}
In the third case, after replacing the second vertex operator in Eq.(\ref{68})
by Eq.(\ref{67}), we get the Wick contraction
\begin{equation}
\int_{0}^{1}dx\frac{1}{M}\left[  -\frac{(e^{T}\cdot k_{4})(2,3)}{(1-x)^{2}%
}\right]  \frac{1}{x}x^{(1,2)}(1-x)^{(2,3)}. \label{80}%
\end{equation}
The high-energy limit of this amplitude, after applying the master formula of
saddle-point approximation, is
\begin{align}
&  \frac{2}{M}E^{2}\tau(e^{T}\cdot k_{3})x^{(1,2)-1}(1-x)^{(2,3)-2}\sqrt
{\frac{\pi\tau}{E^{2}(1-\tau)^{3}}}\nonumber\\
&  =-\frac{4}{M}\sqrt{\pi}E^{2}(1-\tau)^{2}x_{0}^{(1,2)}(1-x_{0})^{(2,3)}.
\label{81}%
\end{align}

In conclusion, from these results, Eqs.(\ref{75}),(\ref{79}),(\ref{81}), we
find the ratios between the 4-point amplitudes associated with $(\alpha
_{-1}^{T})(b_{-\frac{1}{2}}^{T})\left\vert 0,k\right\rangle $, $(\alpha
_{-1}^{L})(b_{-\frac{1}{2}}^{L})\left\vert 0,k\right\rangle $, and
$(b_{-\frac{3}{2}}^{L})\left\vert 0,k\right\rangle $ to be $1:\frac{1}{M^{2}%
}:-\frac{1}{M}$, in perfect agreement with Eqs.(\ref{23}),(\ref{27}) and Eq.
(\ref{ratio}).

\subsection{$M^{2} = 4$}

Our previous examples only involve one fermion operator $b_{-\frac{1}{2}}^{T}%
$, $b_{-\frac{1}{2}}^{L}$, $b_{-\frac{3}{2}}^{L}$. Since in the 4-point
functions with the fixed states $V_{1} \rightarrow$ photon, $V_{3,4}
\rightarrow$ tachyons, the maximum fermion number of the second vertex is
three, it is of interest to see the pattern of stringy amplitudes associated
with the next massive vertices at $M^{2} = 4$.

At this mass level, the relevant states and the vertex operators are (in the
-1 ghost picture)
\begin{align}
(b_{-\frac{1}{2}}^{T})(b_{-\frac{1}{2}}^{L})(b_{-\frac{3}{2}}^{L})\left\vert
0,k\right\rangle  &  \Rightarrow\psi^{T}\psi^{L}\partial\psi^{L}e^{-\phi
}e^{ikX},\label{82}\\
(\alpha_{-1}^{T})(\alpha_{-1}^{T})(b_{-\frac{1}{2}}^{T})\left\vert
0,k\right\rangle  &  \Rightarrow\psi^{T}\partial X^{T}\partial X^{T}e^{-\phi
}e^{ikX}. \label{83}%
\end{align}
To calculate 4-point functions, we can fix the first vertex $(V_{1})$ to be a
photon state in the $-1$ ghost picture, Eq.(\ref{63}), and the third and the
fourth vertices to be tachyon state in the $0$ ghost picture, Eq.(\ref{64}).

Since the applications of saddle-point approximation is essentially identical
to previous cases, we simply list the results of our calculations
\begin{align}
&  (b_{-\frac{1}{2}}^{T})(b_{-\frac{1}{2}}^{L})(b_{-\frac{3}{2}}%
^{L})\left\vert 0,k\right\rangle \nonumber\\
&  \Rightarrow\int_{0}^{1}dx_{2}\langle(\psi_{1}^{T_{1}}e^{-\phi_{1}}%
e^{ik_{1}X_{1}})(\psi_{2}^{T_{2}}\psi_{2}^{L_{2}}\partial\psi_{2}^{L_{2}%
}e^{-\phi_{2}}e^{ik_{2}X_{2}})(k_{3\lambda}\psi_{3}^{\lambda}e^{ik_{3}X_{3}%
})(k_{4\sigma}\psi_{4}^{\sigma}e^{ik_{4}X_{4}})\rangle\nonumber\\
&  =\frac{4\sqrt{\pi}}{M^{2}}E^{3}\tau^{-\frac{1}{2}}(1-\tau)^{\frac{7}{2}%
}x_{0}^{(1,2)}(1-x_{0})^{(2,3)}, \label{84}%
\end{align}%
\begin{align}
&  (\alpha_{-1}^{T})(\alpha_{-1}^{T})(b_{-\frac{1}{2}}^{T})\left\vert
0,k\right\rangle \nonumber\\
&  \Rightarrow\int_{0}^{1}dx_{2}\langle(\psi_{1}^{T_{1}}e^{-\phi_{1}}%
e^{ik_{1}X_{1}})(\psi_{2}^{T_{2}}X_{2}^{L_{2}}X_{2}^{L_{2}}e^{-\phi_{2}%
}e^{ik_{2}X_{2}})(k_{3\lambda}\psi_{3}^{\lambda}e^{ik_{3}X_{3}})(k_{4\sigma
}\psi_{4}^{\sigma}e^{ik_{4}X_{4}})\rangle\nonumber\\
&  =8\sqrt{\pi}E^{3}\tau^{-\frac{1}{2}}(1-\tau)^{\frac{7}{2}}x_{0}%
^{(1,2)}(1-x_{0})^{(2,3)}. \label{85}%
\end{align}
Combining these results, we conclude that the ratio between the $M^{2}=4$
vertices is given by
\begin{equation}
(b_{-\frac{1}{2}}^{T})(b_{-\frac{1}{2}}^{L})(b_{-\frac{3}{2}}^{L})\left\vert
0,k\right\rangle :(\alpha_{-1}^{T})(\alpha_{-1}^{T})(b_{-\frac{1}{2}}%
^{L})\left\vert 0,k\right\rangle =\frac{1}{M^{2}}:2=1:8. \label{86}%
\end{equation}

\subsection{GSO odd vertices at $M^{2}=5$}

In addition to the stringy amplitudes associated with GSO even vertices we
have calculated in the previous subsections, we can also apply the same method
to those associated with the GSO odd vertices. While it is a common practice
to project out the GSO odd states in order to maintain spacetime
supersymmetry, it turns out that we do find linear relation among these
amplitudes. This seems to suggest a hidden structure of superstring theory in
the high-energy limit.

To see this, we examine the vertices of odd GSO parity, at the mass level
$M^{2}=5$. Based on the power-counting rule as in the bosonic string case, we
can identify the relevant vertices and the associated vertex operators as
follows
\begin{align}
(\alpha_{-1}^{T})(b_{-\frac{1}{2}}^{T})(b_{-\frac{3}{2}}^{L})\left\vert
0,k\right\rangle  &  \Rightarrow\psi^{T}\partial\psi^{L}\partial X^{T}%
e^{-\phi}e^{ikX},\label{87}\\
(\alpha_{-1}^{L})(b_{-\frac{1}{2}}^{L})(b_{-\frac{3}{2}}^{L})\left\vert
0,k\right\rangle  &  \Rightarrow\psi^{L}\partial\psi^{L}\partial X^{L}%
e^{-\phi}e^{ikX}. \label{88}%
\end{align}
To calculate 4-point functions, we can fix the first vertex $(V_{1})$ to be a
tachyon state in the $-1$ ghost picture,
\begin{equation}
V_{1}=e^{-\phi_{1}}e^{ik_{1}\cdot X_{1}}, \label{89}%
\end{equation}
and the third and the fourth vertices to be tachyon state in the $0$ ghost
picture, as Eq.(\ref{64}).

Since the applications of saddle-point approximation is essentially identical
to previous cases, we simply list the results of our calculations
\begin{align}
&  (\alpha_{-1}^{T})(b_{-\frac{1}{2}}^{T})(b_{-\frac{3}{2}}^{L})\left\vert
0,k\right\rangle \nonumber\\
&  \Rightarrow\int_{0}^{1}dx_{2}\langle(e^{-\phi_{1}}e^{ik_{1}X_{1}})(\psi
_{2}^{T_{2}}\partial\psi_{2}^{L_{2}}\partial X_{2}^{T_{2}}e^{-\phi_{2}%
}e^{ik_{2}X_{2}})(k_{3\lambda}\psi_{3}^{\lambda}e^{ik_{3}X_{3}})(k_{4\sigma
}\psi_{4}^{\sigma}e^{ik_{4}X_{4}})\rangle,\nonumber\\
&  =-\frac{8\sqrt{\pi}}{M}E^{3}\tau^{-\frac{1}{2}}(1-\tau)^{\frac{7}{2}}%
x_{0}^{(1,2)}(1-x_{0})^{(2,3)}, \label{90}%
\end{align}%
\begin{align}
&  (\alpha_{-1}^{L})(b_{-\frac{1}{2}}^{L})(b_{-\frac{3}{2}}^{L})\left\vert
0,k\right\rangle \nonumber\\
&  \Rightarrow\int_{0}^{1}dx_{2}\langle(e^{-\phi_{1}}e^{ik_{1}X_{1}})(\psi
_{2}^{L_{2}}\partial\psi_{2}^{L_{2}}\partial X_{2}^{L_{2}}e^{-\phi_{2}%
}e^{ik_{2}X_{2}})(k_{3\lambda}\psi_{3}^{\lambda}e^{ik_{3}X_{3}})(k_{4\sigma
}\psi_{4}^{\sigma}e^{ik_{4}X_{4}})\rangle,\nonumber\\
&  =-\frac{4\sqrt{\pi}}{M^{3}}E^{3}\tau^{-\frac{1}{2}}(1-\tau)^{\frac{7}{2}%
}x_{0}^{(1,2)}(1-x_{0})^{(2,3)}. \label{91}%
\end{align}
It is worth noting that in the second calculations, we need to include both
$u^{\prime\prime}(x_{0})$ and $u^{(3)}(x_{0})$ terms of the first order
corrections in saddle-point approximation, Eq.(\ref{70}), to get the correct answer.

Combining these results, we conclude that the ratio between the $M^{2}=3$
vertices is given by
\begin{equation}
(\alpha_{-1}^{T})(b_{-\frac{1}{2}}^{T})(b_{-\frac{3}{2}}^{L})\left\vert
0,k\right\rangle :(\alpha_{-1}^{L})(b_{-\frac{1}{2}}^{L})(b_{-\frac{3}{2}}%
^{L})\left\vert 0,k\right\rangle =2M^{2}:1=10:1. \label{92}%
\end{equation}
Notice that here we also find an interesting connection between GSO even
$M^{2}=4$ amplitudes and those of GSO odd parity at $M^{2}=5$. The high-energy
limits of the four amplitudes, Eqs.(\ref{84}),(\ref{85}),(\ref{90}%
),(\ref{91}), are proportional to each other. and their ratios are $\sqrt
{5}:8\sqrt{5}:(-8):-\frac{4}{5}$.

\section{Polarizations Orthogonal to the Scattering Plane}

In this section we consider high-energy scattering amplitudes of string states
with polarizations $e_{T^{i}},i=3,4...,25,$orthogonal to the scattering plane.
We will present some examples with saddle-point calculations and compare them
with those calculated in section IV. We will find that they are all
proportional to the previous ones considered before. These scattering
amplitudes are of subleading order in energy for the case of 26D open bosonic
string theory. The existence of these new high-energy scattering amplitudes is
due to the worldsheet fermion exchange in the correlation functions as we will
see in the following examples. Our first example is to consider Eq.(\ref{68})
and replace $\psi_{1}^{T_{1}}$ and $\psi_{2}^{T_{2}}$by $\psi_{1}^{T_{1}^{i}}$
and $\psi_{2}^{T_{2}^{i}}$ respectively%

\begin{equation}
\int_{0}^{1}dx_{2}\langle(\psi_{1}^{T_{1}^{i}}e^{-\phi_{1}}e^{ik_{1}X_{1}%
})(\psi_{2}^{T_{2}^{i}}\partial X_{2}^{T_{2}}e^{-\phi_{2}}e^{ik_{2}X_{2}%
})(k_{3\lambda}\psi_{3}^{\lambda}e^{ik_{3}X_{3}})(k_{4\sigma}\psi_{4}^{\sigma
}e^{ik_{4}X_{4}})\rangle. \label{93}%
\end{equation}
The calculation of Eq.(\ref{93}) is similar to that of Eq.(\ref{68}) except
that, for this new case, one ends up with only the first term in
Eq.(\ref{69}), and the second and the third terms vanish. Remarkably, the
final answer is%

\begin{align}
&  -2E^{2}(1-\tau)(e^{T}\cdot k_{3})x_{0}^{(1,2)-1}(1-x_{0})^{(2,3)-1}%
\sqrt{\frac{\pi\tau}{E^{2}(1-\tau)^{3}}}\nonumber\\
&  =-4\sqrt{\pi}E^{2}(1-\tau)^{2}x_{0}^{(1,2)}(1-x_{0})^{(2,3)}, \label{94}%
\end{align}
which is proportional to Eq.(\ref{75}). Our second example is again to replace
$\psi_{1}^{T_{1}}$ and $\psi_{2}^{T_{2}}$ in Eq.(\ref{84}) by $\psi_{1}%
^{T_{1}^{i}}$ and $\psi_{2}^{T_{2}^{i}}$ respectively. One gets exactly the
same answer as Eq.(\ref{84}). The two examples above seem to suggest that
high-energy scattering of string states with polarizations $e_{T^{i}}$ are the
same as that of polarization $e_{T}$ up to a sign. Let's consider the third
example to justify this point. It is straightward to show the following%

\begin{align}
&  \int_{0}^{1}dx_{2}\langle(\psi_{1}^{L}\psi_{1}^{T_{1}}\psi_{1}^{T_{1}^{i}%
}e^{-\phi_{1}}e^{ik_{1}X_{1}})(\psi_{2}^{L}\psi_{2}^{T_{2}}\psi_{2}^{T_{2}%
^{i}}\partial X_{2}^{T_{2}}e^{-\phi_{2}}e^{ik_{2}X_{2}})(k_{3\lambda}\psi
_{3}^{\lambda}e^{ik_{3}X_{3}})(k_{4\sigma}\psi_{4}^{\sigma}e^{ik_{4}X_{4}%
})\rangle\nonumber\\
&  =N[4E^{4}(1-\tau)-4E^{4}(1-\tau)^{2}-4E^{4}\tau(1-\tau)]=0 \label{95}%
\end{align}
On the other hand, if we assume the symmetry for all transverse polarization
vectors $T,T^{i}$ in the scattering amplitudes, one can easily derive the same
conclusion without detailed calculations. Since replacing $T^{i}$ polarization
vectors of both vertices in Eq.(\ref{95}) by $T$ will naturally leads to a
null result due to anti-commuting property of fermions.

It is clear from the above calculations that the existence of these new
high-energy scattering amplitudes of string states with polarizations
$e_{T^{i}}$ orthogonal to the scattering plane is due to the worldsheet
fermion exchange in the correlation functions. These fermion exchanges do not
exist in the pure bosonic string correlation functions and is, presumably,
related to the high-energy massive spacetime fermionic scattering amplitudes
in the R-sector of the theory. Physically, the high-energy scattering
amplitudes of spacetime fermion will enjoy the symmetry of rotations among
different polarizations in the spin space and our results here seem to justify
this observation. If this conjecture turns out to be true, then the list of
vertices we considered in Eqs.(18)-(21) for high-energy stringy amplitudes
should be extended and includes the cases with $b_{-\frac{1}{2}}^{T}$ replaced
by $b_{-\frac{1}{2}}^{T^{i}}$. Obviously, these new high-energy amplitudes
create complications and textures for a full understanding of stringy
symmetry. Nevertheless, the claim that there is only one independent
high-energy scattering amplitude at each fixed mass level of the string
spectrum persists in the case of superstring theory, at least, for the NS
sector of the theory.

\section{Conclusion}

In this paper we have explicitly calculated all high-energy scattering
amplitudes of string states with polarizations on the scattering plane of open
superstring theory. In particular, the proportionality constants among
high-energy scattering amplitudes of different string states at each fixed but
arbitrary mass level are determined by using three different methods. These
constants are shown to originate from zero-norm states in the spectrum as in
the case of open bosonic string theory. In addition, we discover new
high-energy scattering amplitudes, which are still proportional to the
previous ones, with polarizations \textit{orthogonal} to the scattering plane.
We conjecture the existence of a symmetry among high-energy scattering
amplitudes with polarizations $e_{T^{i}}$ and $e_{T}$. These scattering
amplitudes are subleading order in energy for the case of open bosonic string
theory. The existence of these new high-energy scattering amplitudes is due to
the worldsheet fermion exchange in the correlation functions and is argued to
be related to the high-energy massive spacetime fermionic scattering
amplitudes in the R-sector of the theory. Finally, our study also suggests
that the nature of GSO projection in superstring theory might be simplified in
the high-energy limit. Hopefully, this is in connection with the conjecture
that supersymmetry is realized in broken phase without GSO projection in the
open string theory \cite{GSO}.

It would be of crucial importance to calculate high-energy massive fermion
scattering amplitudes in the R-sector to complete the proof of Gross's two
conjectures on high-energy symmetry of superstring theory. The construction of
general \textit{massive} spacetime fermion vertex, involving picture changing,
will be the first step toward understanding of the high-energy behavior of
superstring theory.

\begin{acknowledgments}
The authors thank Yuri Aisaka, Peiming Ho, Yoichi Kazama, Yoshihisa Kitazawa,
Nicolas Moeller, Ryu Sasaki, Shunsuke Teraguch, Peter West, Tamiaki Yoneya for
helpful discussions. This work is supported in part by the National Science
Council and the National Center for Theoretical Sciences, Taiwan, R.O.C.
\end{acknowledgments}

%

\appendix
%

\setcounter{equation}{0}
\renewcommand{\theequation}{\thesection.\arabic{equation}}%

\section{Solve the Virasoro conditions in the high-energy limit}

Let repeat the Virasoro conditions on the general state at the mass level
$M^{2}=\left(  2n-1\right)  $,%
\begin{align}
G_{1/2}\left\vert n\right\rangle  &  =\sum_{m_{j}}\left[  k^{\nu_{1}^{1/2}}%
\begin{tabular}
[c]{|c|c|c|}\hline
$\nu_{1}^{1/2}$ & $\cdots$ & $\nu_{m_{1/2}}^{1/2}$\\\hline
\end{tabular}
^{T}\right.  \overset{k}{\underset{j=1}{\otimes}}%
\begin{tabular}
[c]{|c|c|c|}\hline
$\mu_{1}^{j}$ & $\cdots$ & $\mu_{m_{j}}^{j}$\\\hline
\end{tabular}
\overset{s}{\underset{r\neq1/2}{\otimes}}%
\begin{tabular}
[c]{|c|c|c|}\hline
$\nu_{1}^{r}$ & $\cdots$ & $\nu_{m_{r}}^{r}$\\\hline
\end{tabular}
^{T}\nonumber\\
&  +\sum_{l\geq1}\sum_{i=1}^{m_{l}}l%
\begin{tabular}
[c]{|c|c|c|c|c|}\hline
$\mu_{1}^{l}$ & $\cdots$ & $\hat{\mu}_{i}^{l}$ & $\cdots$ & $\mu_{m_{l}}^{l}%
$\\\hline
\end{tabular}
\otimes%
\begin{tabular}
[c]{|c|c|c|c|}\hline
$\mu_{i}^{l}$ & $\nu_{1}^{l+1/2}$ & $\cdots$ & $\nu_{m_{l+1/2}}^{l+1/2}%
$\\\hline
\end{tabular}
^{T}\nonumber\\
&  \otimes%
\begin{tabular}
[c]{|c|c|c|}\hline
$\nu_{2}^{1/2}$ & $\cdots$ & $\nu_{m_{1/2}}^{1/2}$\\\hline
\end{tabular}
^{T}\overset{k}{\underset{j\neq l}{\otimes}}%
\begin{tabular}
[c]{|c|c|c|}\hline
$\mu_{1}^{j}$ & $\cdots$ & $\mu_{m_{j}}^{j}$\\\hline
\end{tabular}
\overset{s}{\underset{r\neq1/2,l+1/2}{\otimes}}%
\begin{tabular}
[c]{|c|c|c|}\hline
$\nu_{1}^{r}$ & $\cdots$ & $\nu_{m_{r}}^{r}$\\\hline
\end{tabular}
^{T}\nonumber\\
&  +\sum_{i=2}^{m_{1/2}}%
\begin{tabular}
[c]{|c|c|c|c|}\hline
$\nu_{i}^{1/2}$ & $\mu_{1}^{1}$ & $\cdots$ & $\mu_{m_{1}}^{1}$\\\hline
\end{tabular}
\otimes\left(  -1\right)  ^{i+1}%
\begin{tabular}
[c]{|c|c|c|c|c|}\hline
$\nu_{2}^{1/2}$ & $\cdots$ & $\hat{\nu}_{i}^{1/2}$ & $\cdots$ & $\nu_{m_{1/2}%
}^{1/2}$\\\hline
\end{tabular}
^{T}\nonumber\\
&  \overset{k}{\underset{j\neq l}{\otimes}}%
\begin{tabular}
[c]{|c|c|c|}\hline
$\mu_{1}^{j}$ & $\cdots$ & $\mu_{m_{j}}^{j}$\\\hline
\end{tabular}
\overset{s}{\underset{r\neq1/2}{\otimes}}%
\begin{tabular}
[c]{|c|c|c|}\hline
$\nu_{1}^{r}$ & $\cdots$ & $\nu_{m_{r}}^{r}$\\\hline
\end{tabular}
^{T}\nonumber\\
&  +\sum_{l\geq2}\sum_{i=1}^{m_{l-1/2}}\left(  -1\right)  ^{i+1}%
\begin{tabular}
[c]{|c|c|c|c|}\hline
$\nu_{i}^{l-1/2}$ & $\mu_{1}^{l}$ & $\cdots$ & $\mu_{m_{l}}^{l}$\\\hline
\end{tabular}
\otimes%
\begin{tabular}
[c]{|c|c|c|c|c|}\hline
$\nu_{1}^{l-1/2}$ & $\cdots$ & $\hat{\nu}_{i}^{l-1/2}$ & $\cdots$ &
$\nu_{m_{l-1/2}}^{l-1/2}$\\\hline
\end{tabular}
^{T}\nonumber\\
&  \otimes%
\begin{tabular}
[c]{|c|c|c|}\hline
$\nu_{2}^{1/2}$ & $\cdots$ & $\nu_{m_{1/2}}^{1/2}$\\\hline
\end{tabular}
^{T}\left.  \overset{k}{\underset{j\neq l}{\otimes}}%
\begin{tabular}
[c]{|c|c|c|}\hline
$\mu_{1}^{j}$ & $\cdots$ & $\mu_{m_{j}}^{j}$\\\hline
\end{tabular}
\overset{s}{\underset{r\neq1/2,l-1/2}{\otimes}}%
\begin{tabular}
[c]{|c|c|c|}\hline
$\nu_{1}^{r}$ & $\cdots$ & $\nu_{m_{r}}^{r}$\\\hline
\end{tabular}
^{T}\right] \nonumber\\
&  \cdot\frac{1}{\left(  m_{1/2}-1\right)  !}\psi_{-1/2}^{\nu_{2}^{1/2}%
\cdots\nu_{m_{1/2}}^{1/2}}\prod_{j=1}^{k}\frac{1}{j^{m_{j}}m_{j}!}\alpha
_{-j}^{\mu_{1}^{j}\cdots\mu_{m_{j}}^{j}}\prod_{r\neq1/2}^{s}\frac{1}{m_{r}%
!}\psi_{-r}^{\nu_{1}^{r}\cdots\nu_{m_{r}}^{r}},
\end{align}
and%
\begin{align}
G_{3/2}\left\vert n\right\rangle  &  =\sum_{m_{j}}\left[
\begin{tabular}
[c]{|c|c|c|}\hline
$\nu_{1}^{3/2}$ & $\cdots$ & $\nu_{m_{3/2}}^{3/2}$\\\hline
\end{tabular}
^{T}k^{\nu_{1}^{3/2}}\right.  \overset{k}{\underset{j=1}{\otimes}}%
\begin{tabular}
[c]{|c|c|c|}\hline
$\mu_{1}^{j}$ & $\cdots$ & $\mu_{m_{j}}^{j}$\\\hline
\end{tabular}
\overset{s}{\underset{r\neq3/2}{\otimes}}%
\begin{tabular}
[c]{|c|c|c|}\hline
$\nu_{1}^{r}$ & $\cdots$ & $\nu_{m_{r}}^{r}$\\\hline
\end{tabular}
^{T}\nonumber\\
&  +\eta^{\mu\nu}%
\begin{tabular}
[c]{|c|c|c|c|}\hline
$\mu$ & $\mu_{1}^{1}$ & $\cdots$ & $\mu_{m_{1}}^{1}$\\\hline
\end{tabular}
\otimes%
\begin{tabular}
[c]{|c|c|c|c|}\hline
$\nu$ & $\nu_{1}^{1/2}$ & $\cdots$ & $\nu_{m_{1/2}}^{1/2}$\\\hline
\end{tabular}
^{T}\otimes%
\begin{tabular}
[c]{|c|c|c|}\hline
$\nu_{2}^{3/2}$ & $\cdots$ & $\nu_{m_{3/2}}^{3/2}$\\\hline
\end{tabular}
^{T}\nonumber\\
&  \overset{k}{\underset{j\neq1}{\otimes}}%
\begin{tabular}
[c]{|c|c|c|}\hline
$\mu_{1}^{j}$ & $\cdots$ & $\mu_{m_{j}}^{j}$\\\hline
\end{tabular}
\overset{s}{\underset{r\neq1/2,3/2}{\otimes}}%
\begin{tabular}
[c]{|c|c|c|}\hline
$\nu_{1}^{r}$ & $\cdots$ & $\nu_{m_{r}}^{r}$\\\hline
\end{tabular}
^{T}\nonumber\\
&  +\sum_{l\geq1}\sum_{i=1}^{m_{l}}l%
\begin{tabular}
[c]{|c|c|c|c|c|}\hline
$\mu_{1}^{l}$ & $\cdots$ & $\hat{\mu}_{i}^{l}$ & $\cdots$ & $\mu_{m_{l}}^{l}%
$\\\hline
\end{tabular}
\otimes%
\begin{tabular}
[c]{|c|c|c|c|}\hline
$\mu_{i}^{l}$ & $\nu_{1}^{l+3/2}$ & $\cdots$ & $\nu_{m_{l+3/2}}^{l+3/2}%
$\\\hline
\end{tabular}
^{T}\nonumber\\
&  \otimes%
\begin{tabular}
[c]{|c|c|c|}\hline
$\nu_{2}^{3/2}$ & $\cdots$ & $\nu_{m_{3/2}}^{3/2}$\\\hline
\end{tabular}
^{T}\overset{k}{\underset{j\neq l}{\otimes}}%
\begin{tabular}
[c]{|c|c|c|}\hline
$\mu_{1}^{j}$ & $\cdots$ & $\mu_{m_{j}}^{j}$\\\hline
\end{tabular}
\overset{s}{\underset{r\neq3/2,l+3/2}{\otimes}}%
\begin{tabular}
[c]{|c|c|c|}\hline
$\nu_{1}^{r}$ & $\cdots$ & $\nu_{m_{r}}^{r}$\\\hline
\end{tabular}
^{T}\nonumber\\
&  +\sum_{i=2}^{m_{3/2}}3%
\begin{tabular}
[c]{|c|c|c|c|}\hline
$\nu_{i}^{3/2}$ & $\mu_{1}^{3}$ & $\cdots$ & $\mu_{m_{3}}^{3}$\\\hline
\end{tabular}
\otimes\left(  -1\right)  ^{i+1}%
\begin{tabular}
[c]{|c|c|c|c|c|}\hline
$\nu_{2}^{3/2}$ & $\cdots$ & $\hat{\nu}_{i}^{3/2}$ & $\cdots$ & $\nu_{m_{3/2}%
}^{3/2}$\\\hline
\end{tabular}
^{T}\nonumber\\
&  \overset{k}{\underset{j\neq3}{\otimes}}%
\begin{tabular}
[c]{|c|c|c|}\hline
$\mu_{1}^{j}$ & $\cdots$ & $\mu_{m_{j}}^{j}$\\\hline
\end{tabular}
\overset{s}{\underset{r\neq3/2}{\otimes}}%
\begin{tabular}
[c]{|c|c|c|}\hline
$\nu_{1}^{r}$ & $\cdots$ & $\nu_{m_{r}}^{r}$\\\hline
\end{tabular}
^{T}\nonumber\\
&  +\sum_{l\geq2,l\neq3}\sum_{i=1}^{m_{l-3/2}}%
\begin{tabular}
[c]{|c|c|c|c|}\hline
$\nu_{i}^{l-3/2}$ & $\mu_{1}^{l}$ & $\cdots$ & $\mu_{m_{l}}^{l}$\\\hline
\end{tabular}
\otimes%
\begin{tabular}
[c]{|c|c|c|c|c|}\hline
$\nu_{1}^{l-3/2}$ & $\cdots$ & $\hat{\nu}_{i}^{l-3/2}$ & $\cdots$ &
$\nu_{m_{l-3/2}}^{l-3/2}$\\\hline
\end{tabular}
^{T}\nonumber\\
&  \otimes%
\begin{tabular}
[c]{|c|c|c|}\hline
$\nu_{2}^{3/2}$ & $\cdots$ & $\nu_{m_{3/2}}^{3/2}$\\\hline
\end{tabular}
^{T}\left.  \overset{k}{\underset{j\neq l}{\otimes}}%
\begin{tabular}
[c]{|c|c|c|}\hline
$\mu_{1}^{j}$ & $\cdots$ & $\mu_{m_{j}}^{j}$\\\hline
\end{tabular}
\overset{s}{\underset{r\neq3/2,l-3/2}{\otimes}}%
\begin{tabular}
[c]{|c|c|c|}\hline
$\nu_{1}^{r}$ & $\cdots$ & $\nu_{m_{r}}^{r}$\\\hline
\end{tabular}
^{T}\right] \nonumber\\
&  \cdot\frac{1}{\left(  m_{3/2}-1\right)  !}\psi_{-3/2}^{\nu_{2}^{3/2}%
\cdots\nu_{m_{3/2}}^{3/2}}\prod_{j=1}^{k}\frac{1}{j^{m_{j}}m_{j}!}\alpha
_{-j}^{\mu_{1}^{j}\cdots\mu_{m_{j}}^{j}}\prod_{r\neq3/2}^{s}\frac{1}{m_{r}%
!}\psi_{-r}^{\nu_{1}^{r}\cdots\nu_{m_{r}}^{r}}.
\end{align}

We then obtain the constrain equations%
\begin{subequations}%
\begin{align}
0  &  =k^{\nu_{1}^{1/2}}%
\begin{tabular}
[c]{|c|c|c|}\hline
$\nu_{1}^{1/2}$ & $\cdots$ & $\nu_{m_{1/2}}^{1/2}$\\\hline
\end{tabular}
^{T}\overset{k}{\underset{j=1}{\otimes}}%
\begin{tabular}
[c]{|c|c|c|}\hline
$\mu_{1}^{j}$ & $\cdots$ & $\mu_{m_{j}}^{j}$\\\hline
\end{tabular}
\overset{s}{\underset{r\neq1/2}{\otimes}}%
\begin{tabular}
[c]{|c|c|c|}\hline
$\nu_{1}^{r}$ & $\cdots$ & $\nu_{m_{r}}^{r}$\\\hline
\end{tabular}
^{T}\nonumber\\
&  +\sum_{l\geq1}\sum_{i=1}^{m_{l}}l%
\begin{tabular}
[c]{|c|c|c|c|c|}\hline
$\mu_{1}^{l}$ & $\cdots$ & $\hat{\mu}_{i}^{l}$ & $\cdots$ & $\mu_{m_{l}}^{l}%
$\\\hline
\end{tabular}
\otimes%
\begin{tabular}
[c]{|c|c|c|c|}\hline
$\mu_{i}^{l}$ & $\nu_{1}^{l+1/2}$ & $\cdots$ & $\nu_{m_{l+1/2}}^{l+1/2}%
$\\\hline
\end{tabular}
^{T}\nonumber\\
&  \otimes%
\begin{tabular}
[c]{|c|c|c|}\hline
$\nu_{2}^{1/2}$ & $\cdots$ & $\nu_{m_{1/2}}^{1/2}$\\\hline
\end{tabular}
^{T}\overset{k}{\underset{j\neq l}{\otimes}}%
\begin{tabular}
[c]{|c|c|c|}\hline
$\mu_{1}^{j}$ & $\cdots$ & $\mu_{m_{j}}^{j}$\\\hline
\end{tabular}
\overset{s}{\underset{r\neq1/2,l+1/2}{\otimes}}%
\begin{tabular}
[c]{|c|c|c|}\hline
$\nu_{1}^{r}$ & $\cdots$ & $\nu_{m_{r}}^{r}$\\\hline
\end{tabular}
^{T}\nonumber\\
&  +\sum_{i=2}^{m_{1/2}}%
\begin{tabular}
[c]{|c|c|c|c|}\hline
$\nu_{i}^{1/2}$ & $\mu_{1}^{1}$ & $\cdots$ & $\mu_{m_{1}}^{1}$\\\hline
\end{tabular}
\otimes\left(  -1\right)  ^{i+1}%
\begin{tabular}
[c]{|c|c|c|c|c|}\hline
$\nu_{2}^{1/2}$ & $\cdots$ & $\hat{\nu}_{i}^{1/2}$ & $\cdots$ & $\nu_{m_{1/2}%
}^{1/2}$\\\hline
\end{tabular}
^{T}\nonumber\\
&  \overset{k}{\underset{j\neq l}{\otimes}}%
\begin{tabular}
[c]{|c|c|c|}\hline
$\mu_{1}^{j}$ & $\cdots$ & $\mu_{m_{j}}^{j}$\\\hline
\end{tabular}
\overset{s}{\underset{r\neq1/2}{\otimes}}%
\begin{tabular}
[c]{|c|c|c|}\hline
$\nu_{1}^{r}$ & $\cdots$ & $\nu_{m_{r}}^{r}$\\\hline
\end{tabular}
^{T}\nonumber\\
&  +\sum_{l\geq2}\sum_{i=1}^{m_{l-1/2}}\left(  -1\right)  ^{i+1}%
\begin{tabular}
[c]{|c|c|c|c|}\hline
$\nu_{i}^{l-1/2}$ & $\mu_{1}^{l}$ & $\cdots$ & $\mu_{m_{l}}^{l}$\\\hline
\end{tabular}
\otimes%
\begin{tabular}
[c]{|c|c|c|c|c|}\hline
$\nu_{1}^{l-1/2}$ & $\cdots$ & $\hat{\nu}_{i}^{l-1/2}$ & $\cdots$ &
$\nu_{m_{l-1/2}}^{l-1/2}$\\\hline
\end{tabular}
^{T}\nonumber\\
&  \otimes%
\begin{tabular}
[c]{|c|c|c|}\hline
$\nu_{2}^{1/2}$ & $\cdots$ & $\nu_{m_{1/2}}^{1/2}$\\\hline
\end{tabular}
^{T}\overset{k}{\underset{j\neq l}{\otimes}}%
\begin{tabular}
[c]{|c|c|c|}\hline
$\mu_{1}^{j}$ & $\cdots$ & $\mu_{m_{j}}^{j}$\\\hline
\end{tabular}
\overset{s}{\underset{r\neq1/2,l-1/2}{\otimes}}%
\begin{tabular}
[c]{|c|c|c|}\hline
$\nu_{1}^{r}$ & $\cdots$ & $\nu_{m_{r}}^{r}$\\\hline
\end{tabular}
^{T},
\end{align}%
\begin{align}
0  &  =%
\begin{tabular}
[c]{|c|c|c|}\hline
$\nu_{1}^{3/2}$ & $\cdots$ & $\nu_{m_{3/2}}^{3/2}$\\\hline
\end{tabular}
^{T}k^{\nu_{1}^{3/2}}\overset{k}{\underset{j=1}{\otimes}}%
\begin{tabular}
[c]{|c|c|c|}\hline
$\mu_{1}^{j}$ & $\cdots$ & $\mu_{m_{j}}^{j}$\\\hline
\end{tabular}
\overset{s}{\underset{r\neq3/2}{\otimes}}%
\begin{tabular}
[c]{|c|c|c|}\hline
$\nu_{1}^{r}$ & $\cdots$ & $\nu_{m_{r}}^{r}$\\\hline
\end{tabular}
^{T}\nonumber\\
&  +\eta^{\mu\nu}%
\begin{tabular}
[c]{|c|c|c|c|}\hline
$\mu$ & $\mu_{1}^{1}$ & $\cdots$ & $\mu_{m_{1}}^{1}$\\\hline
\end{tabular}
\otimes%
\begin{tabular}
[c]{|c|c|c|c|}\hline
$\nu$ & $\nu_{1}^{1/2}$ & $\cdots$ & $\nu_{m_{1/2}}^{1/2}$\\\hline
\end{tabular}
^{T}\otimes%
\begin{tabular}
[c]{|c|c|c|}\hline
$\nu_{2}^{3/2}$ & $\cdots$ & $\nu_{m_{3/2}}^{3/2}$\\\hline
\end{tabular}
^{T}\nonumber\\
&  \overset{k}{\underset{j\neq1}{\otimes}}%
\begin{tabular}
[c]{|c|c|c|}\hline
$\mu_{1}^{j}$ & $\cdots$ & $\mu_{m_{j}}^{j}$\\\hline
\end{tabular}
\overset{s}{\underset{r\neq1/2,3/2}{\otimes}}%
\begin{tabular}
[c]{|c|c|c|}\hline
$\nu_{1}^{r}$ & $\cdots$ & $\nu_{m_{r}}^{r}$\\\hline
\end{tabular}
^{T}\nonumber\\
&  +\sum_{l\geq1}\sum_{i=1}^{m_{l}}l%
\begin{tabular}
[c]{|c|c|c|c|c|}\hline
$\mu_{1}^{l}$ & $\cdots$ & $\hat{\mu}_{i}^{l}$ & $\cdots$ & $\mu_{m_{l}}^{l}%
$\\\hline
\end{tabular}
\otimes%
\begin{tabular}
[c]{|c|c|c|c|}\hline
$\mu_{i}^{l}$ & $\nu_{1}^{l+3/2}$ & $\cdots$ & $\nu_{m_{l+3/2}}^{l+3/2}%
$\\\hline
\end{tabular}
^{T}\nonumber\\
&  \otimes%
\begin{tabular}
[c]{|c|c|c|}\hline
$\nu_{2}^{3/2}$ & $\cdots$ & $\nu_{m_{3/2}}^{3/2}$\\\hline
\end{tabular}
^{T}\overset{k}{\underset{j\neq l}{\otimes}}%
\begin{tabular}
[c]{|c|c|c|}\hline
$\mu_{1}^{j}$ & $\cdots$ & $\mu_{m_{j}}^{j}$\\\hline
\end{tabular}
\overset{s}{\underset{r\neq3/2,l+3/2}{\otimes}}%
\begin{tabular}
[c]{|c|c|c|}\hline
$\nu_{1}^{r}$ & $\cdots$ & $\nu_{m_{r}}^{r}$\\\hline
\end{tabular}
^{T}\nonumber\\
&  +\sum_{i=2}^{m_{3/2}}3%
\begin{tabular}
[c]{|c|c|c|c|}\hline
$\nu_{i}^{3/2}$ & $\mu_{1}^{3}$ & $\cdots$ & $\mu_{m_{3}}^{3}$\\\hline
\end{tabular}
\otimes\left(  -1\right)  ^{i+1}%
\begin{tabular}
[c]{|c|c|c|c|c|}\hline
$\nu_{2}^{3/2}$ & $\cdots$ & $\hat{\nu}_{i}^{3/2}$ & $\cdots$ & $\nu_{m_{3/2}%
}^{3/2}$\\\hline
\end{tabular}
^{T}\nonumber\\
&  \overset{k}{\underset{j\neq3}{\otimes}}%
\begin{tabular}
[c]{|c|c|c|}\hline
$\mu_{1}^{j}$ & $\cdots$ & $\mu_{m_{j}}^{j}$\\\hline
\end{tabular}
\overset{s}{\underset{r\neq3/2}{\otimes}}%
\begin{tabular}
[c]{|c|c|c|}\hline
$\nu_{1}^{r}$ & $\cdots$ & $\nu_{m_{r}}^{r}$\\\hline
\end{tabular}
^{T}\nonumber\\
&  +\sum_{l\geq2,l\neq3}\sum_{i=1}^{m_{l-3/2}}%
\begin{tabular}
[c]{|c|c|c|c|}\hline
$\nu_{i}^{l-3/2}$ & $\mu_{1}^{l}$ & $\cdots$ & $\mu_{m_{l}}^{l}$\\\hline
\end{tabular}
\otimes\left(  -1\right)  ^{i+1}%
\begin{tabular}
[c]{|c|c|c|c|c|}\hline
$\nu_{1}^{l-3/2}$ & $\cdots$ & $\hat{\nu}_{i}^{l-3/2}$ & $\cdots$ &
$\nu_{m_{l-3/2}}^{l-3/2}$\\\hline
\end{tabular}
^{T}\nonumber\\
&  \otimes%
\begin{tabular}
[c]{|c|c|c|}\hline
$\nu_{2}^{3/2}$ & $\cdots$ & $\nu_{m_{3/2}}^{3/2}$\\\hline
\end{tabular}
^{T}\overset{k}{\underset{j\neq l}{\otimes}}%
\begin{tabular}
[c]{|c|c|c|}\hline
$\mu_{1}^{j}$ & $\cdots$ & $\mu_{m_{j}}^{j}$\\\hline
\end{tabular}
\overset{s}{\underset{r\neq3/2,l-3/2}{\otimes}}%
\begin{tabular}
[c]{|c|c|c|}\hline
$\nu_{1}^{r}$ & $\cdots$ & $\nu_{m_{r}}^{r}$\\\hline
\end{tabular}
^{T}.
\end{align}%
\end{subequations}%
Taking the high-energy limit in the above equations by letting $\left(
\mu_{i},\nu_{i}\right)  \rightarrow\left(  L,T\right)  $, and%
\begin{equation}
k^{\mu_{i}}\rightarrow Me^{L}\text{, }\eta^{\mu_{1}\mu_{2}}\rightarrow
e^{T}e^{T}.
\end{equation}
We get%
\begin{subequations}%
\begin{align}
0  &  =M%
\begin{tabular}
[c]{|c|c|c|c|}\hline
$L$ & $\nu_{2}^{1/2}$ & $\cdots$ & $\nu_{m_{1/2}}^{1/2}$\\\hline
\end{tabular}
^{T}\overset{k}{\underset{j=1}{\otimes}}%
\begin{tabular}
[c]{|c|c|c|}\hline
$\mu_{1}^{j}$ & $\cdots$ & $\mu_{m_{j}}^{j}$\\\hline
\end{tabular}
\overset{s}{\underset{r\neq1/2}{\otimes}}%
\begin{tabular}
[c]{|c|c|c|}\hline
$\nu_{1}^{r}$ & $\cdots$ & $\nu_{m_{r}}^{r}$\\\hline
\end{tabular}
^{T}\nonumber\\
&  +\sum_{l\geq1}\sum_{i=1}^{m_{l}}l%
\begin{tabular}
[c]{|c|c|c|c|c|}\hline
$\mu_{1}^{l}$ & $\cdots$ & $\hat{\mu}_{i}^{l}$ & $\cdots$ & $\mu_{m_{l}}^{l}%
$\\\hline
\end{tabular}
\otimes%
\begin{tabular}
[c]{|c|c|c|c|}\hline
$\mu_{i}^{l}$ & $\nu_{1}^{l+1/2}$ & $\cdots$ & $\nu_{m_{l+1/2}}^{l+1/2}%
$\\\hline
\end{tabular}
^{T}\nonumber\\
&  \otimes%
\begin{tabular}
[c]{|c|c|c|}\hline
$\nu_{2}^{1/2}$ & $\cdots$ & $\nu_{m_{1/2}}^{1/2}$\\\hline
\end{tabular}
^{T}\overset{k}{\underset{j\neq l}{\otimes}}%
\begin{tabular}
[c]{|c|c|c|}\hline
$\mu_{1}^{j}$ & $\cdots$ & $\mu_{m_{j}}^{j}$\\\hline
\end{tabular}
\overset{s}{\underset{r\neq1/2,l+1/2}{\otimes}}%
\begin{tabular}
[c]{|c|c|c|}\hline
$\nu_{1}^{r}$ & $\cdots$ & $\nu_{m_{r}}^{r}$\\\hline
\end{tabular}
^{T}\nonumber\\
&  +\sum_{i=2}^{m_{1/2}}%
\begin{tabular}
[c]{|c|c|c|c|}\hline
$\nu_{i}^{1/2}$ & $\mu_{1}^{1}$ & $\cdots$ & $\mu_{m_{1}}^{1}$\\\hline
\end{tabular}
\otimes\left(  -1\right)  ^{i+1}%
\begin{tabular}
[c]{|c|c|c|c|c|}\hline
$\nu_{2}^{1/2}$ & $\cdots$ & $\hat{\nu}_{i}^{1/2}$ & $\cdots$ & $\nu_{m_{1/2}%
}^{1/2}$\\\hline
\end{tabular}
^{T}\nonumber\\
&  \overset{k}{\underset{j\neq l}{\otimes}}%
\begin{tabular}
[c]{|c|c|c|}\hline
$\mu_{1}^{j}$ & $\cdots$ & $\mu_{m_{j}}^{j}$\\\hline
\end{tabular}
\overset{s}{\underset{r\neq1/2}{\otimes}}%
\begin{tabular}
[c]{|c|c|c|}\hline
$\nu_{1}^{r}$ & $\cdots$ & $\nu_{m_{r}}^{r}$\\\hline
\end{tabular}
^{T}\nonumber\\
&  +\sum_{l\geq2}\sum_{i=1}^{m_{l-1/2}}\left(  -1\right)  ^{i+1}%
\begin{tabular}
[c]{|c|c|c|c|}\hline
$\nu_{i}^{l-1/2}$ & $\mu_{1}^{l}$ & $\cdots$ & $\mu_{m_{l}}^{l}$\\\hline
\end{tabular}
\otimes%
\begin{tabular}
[c]{|c|c|c|c|c|}\hline
$\nu_{1}^{l-1/2}$ & $\cdots$ & $\hat{\nu}_{i}^{l-1/2}$ & $\cdots$ &
$\nu_{m_{l-1/2}}^{l-1/2}$\\\hline
\end{tabular}
^{T}\nonumber\\
&  \otimes%
\begin{tabular}
[c]{|c|c|c|}\hline
$\nu_{2}^{1/2}$ & $\cdots$ & $\nu_{m_{1/2}}^{1/2}$\\\hline
\end{tabular}
^{T}\overset{k}{\underset{j\neq l}{\otimes}}%
\begin{tabular}
[c]{|c|c|c|}\hline
$\mu_{1}^{j}$ & $\cdots$ & $\mu_{m_{j}}^{j}$\\\hline
\end{tabular}
\overset{s}{\underset{r\neq1/2,l-1/2}{\otimes}}%
\begin{tabular}
[c]{|c|c|c|}\hline
$\nu_{1}^{r}$ & $\cdots$ & $\nu_{m_{r}}^{r}$\\\hline
\end{tabular}
^{T},
\end{align}%
\begin{align}
0  &  =M%
\begin{tabular}
[c]{|c|c|c|c|}\hline
$L$ & $\nu_{2}^{3/2}$ & $\cdots$ & $\nu_{m_{3/2}}^{3/2}$\\\hline
\end{tabular}
^{T}\overset{k}{\underset{j=1}{\otimes}}%
\begin{tabular}
[c]{|c|c|c|}\hline
$\mu_{1}^{j}$ & $\cdots$ & $\mu_{m_{j}}^{j}$\\\hline
\end{tabular}
\overset{s}{\underset{r\neq3/2}{\otimes}}%
\begin{tabular}
[c]{|c|c|c|}\hline
$\nu_{1}^{r}$ & $\cdots$ & $\nu_{m_{r}}^{r}$\\\hline
\end{tabular}
^{T}\nonumber\\
&  +%
\begin{tabular}
[c]{|c|c|c|c|}\hline
$T$ & $\mu_{1}^{1}$ & $\cdots$ & $\mu_{m_{1}}^{1}$\\\hline
\end{tabular}
\otimes%
\begin{tabular}
[c]{|c|c|c|c|}\hline
$T$ & $\nu_{1}^{1/2}$ & $\cdots$ & $\nu_{m_{1/2}}^{1/2}$\\\hline
\end{tabular}
^{T}\otimes%
\begin{tabular}
[c]{|c|c|c|}\hline
$\nu_{2}^{3/2}$ & $\cdots$ & $\nu_{m_{3/2}}^{3/2}$\\\hline
\end{tabular}
^{T}\nonumber\\
&  \overset{k}{\underset{j\neq1}{\otimes}}%
\begin{tabular}
[c]{|c|c|c|}\hline
$\mu_{1}^{j}$ & $\cdots$ & $\mu_{m_{j}}^{j}$\\\hline
\end{tabular}
\overset{s}{\underset{r\neq1/2,3/2}{\otimes}}%
\begin{tabular}
[c]{|c|c|c|}\hline
$\nu_{1}^{r}$ & $\cdots$ & $\nu_{m_{r}}^{r}$\\\hline
\end{tabular}
^{T}\nonumber\\
&  +\sum_{l\geq1}\sum_{i=1}^{m_{l}}l%
\begin{tabular}
[c]{|c|c|c|c|c|}\hline
$\mu_{1}^{l}$ & $\cdots$ & $\hat{\mu}_{i}^{l}$ & $\cdots$ & $\mu_{m_{l}}^{l}%
$\\\hline
\end{tabular}
\otimes%
\begin{tabular}
[c]{|c|c|c|c|}\hline
$\mu_{i}^{l}$ & $\nu_{1}^{l+3/2}$ & $\cdots$ & $\nu_{m_{l+3/2}}^{l+3/2}%
$\\\hline
\end{tabular}
^{T}\nonumber\\
&  \otimes%
\begin{tabular}
[c]{|c|c|c|}\hline
$\nu_{2}^{3/2}$ & $\cdots$ & $\nu_{m_{3/2}}^{3/2}$\\\hline
\end{tabular}
^{T}\overset{k}{\underset{j\neq l}{\otimes}}%
\begin{tabular}
[c]{|c|c|c|}\hline
$\mu_{1}^{j}$ & $\cdots$ & $\mu_{m_{j}}^{j}$\\\hline
\end{tabular}
\overset{s}{\underset{r\neq3/2,l+3/2}{\otimes}}%
\begin{tabular}
[c]{|c|c|c|}\hline
$\nu_{1}^{r}$ & $\cdots$ & $\nu_{m_{r}}^{r}$\\\hline
\end{tabular}
^{T}\nonumber\\
&  +\sum_{i=2}^{m_{3/2}}3%
\begin{tabular}
[c]{|c|c|c|c|}\hline
$\nu_{i}^{3/2}$ & $\mu_{1}^{3}$ & $\cdots$ & $\mu_{m_{3}}^{3}$\\\hline
\end{tabular}
\otimes\left(  -1\right)  ^{i+1}%
\begin{tabular}
[c]{|c|c|c|c|c|}\hline
$\nu_{2}^{3/2}$ & $\cdots$ & $\hat{\nu}_{i}^{3/2}$ & $\cdots$ & $\nu_{m_{3/2}%
}^{3/2}$\\\hline
\end{tabular}
^{T}\nonumber\\
&  \overset{k}{\underset{j\neq3}{\otimes}}%
\begin{tabular}
[c]{|c|c|c|}\hline
$\mu_{1}^{j}$ & $\cdots$ & $\mu_{m_{j}}^{j}$\\\hline
\end{tabular}
\overset{s}{\underset{r\neq3/2}{\otimes}}%
\begin{tabular}
[c]{|c|c|c|}\hline
$\nu_{1}^{r}$ & $\cdots$ & $\nu_{m_{r}}^{r}$\\\hline
\end{tabular}
^{T}\nonumber\\
&  +\sum_{l\geq2,l\neq3}\sum_{i=1}^{m_{l-3/2}}%
\begin{tabular}
[c]{|c|c|c|c|}\hline
$\nu_{i}^{l-3/2}$ & $\mu_{1}^{l}$ & $\cdots$ & $\mu_{m_{l}}^{l}$\\\hline
\end{tabular}
\otimes\left(  -1\right)  ^{i+1}%
\begin{tabular}
[c]{|c|c|c|c|c|}\hline
$\nu_{1}^{l-3/2}$ & $\cdots$ & $\hat{\nu}_{i}^{l-3/2}$ & $\cdots$ &
$\nu_{m_{l-3/2}}^{l-3/2}$\\\hline
\end{tabular}
^{T}\nonumber\\
&  \otimes%
\begin{tabular}
[c]{|c|c|c|}\hline
$\nu_{2}^{3/2}$ & $\cdots$ & $\nu_{m_{3/2}}^{3/2}$\\\hline
\end{tabular}
^{T}\overset{k}{\underset{j\neq l}{\otimes}}%
\begin{tabular}
[c]{|c|c|c|}\hline
$\mu_{1}^{j}$ & $\cdots$ & $\mu_{m_{j}}^{j}$\\\hline
\end{tabular}
\overset{s}{\underset{r\neq3/2,l-3/2}{\otimes}}%
\begin{tabular}
[c]{|c|c|c|}\hline
$\nu_{1}^{r}$ & $\cdots$ & $\nu_{m_{r}}^{r}$\\\hline
\end{tabular}
^{T}.
\end{align}%
\end{subequations}%
The indices$\left\{  \mu_{i}^{j}\right\}  $ are symmetric and can be chosen to
have $l_{j}$ of $\left\{  L\right\}  $ and $\left\{  T\right\}  $, while
$\left\{  \nu_{i}^{r}\right\}  $ are antisymmetric and we keep them as what
they are at this moment. Thus%
\begin{subequations}%
\begin{align}
0  &  =M%
\begin{tabular}
[c]{|c|c|c|c|}\hline
$L$ & $\nu_{2}^{1/2}$ & $\cdots$ & $\nu_{m_{1/2}}^{1/2}$\\\hline
\end{tabular}
^{T}\overset{k}{\underset{j=1}{\otimes}}%
\begin{tabular}
[c]{|c|}\hline
$\mu_{1}^{j}$\\\hline
\end{tabular}
\underset{m_{j}-1-l_{j}}{\underbrace{%
\begin{tabular}
[c]{|l|l|l|}\hline
$T$ & $\cdots$ & $T$\\\hline
\end{tabular}
}}\underset{l_{j}}{\underbrace{%
\begin{tabular}
[c]{|l|l|l|}\hline
$L$ & $\cdots$ & $L$\\\hline
\end{tabular}
}}\overset{s}{\underset{r\neq1/2}{\otimes}}%
\begin{tabular}
[c]{|c|c|c|}\hline
$\nu_{1}^{r}$ & $\cdots$ & $\nu_{m_{r}}^{r}$\\\hline
\end{tabular}
^{T}\nonumber\\
&  +\sum_{l\geq1}l\underset{m_{l}-1-l_{l}}{\underbrace{%
\begin{tabular}
[c]{|l|l|l|}\hline
$T$ & $\cdots$ & $T$\\\hline
\end{tabular}
}}\underset{l_{l}}{\underbrace{%
\begin{tabular}
[c]{|l|l|l|}\hline
$L$ & $\cdots$ & $L$\\\hline
\end{tabular}
}}\otimes%
\begin{tabular}
[c]{|c|c|c|c|}\hline
$\mu_{1}^{l}$ & $\nu_{1}^{l+1/2}$ & $\cdots$ & $\nu_{m_{l+1/2}}^{l+1/2}%
$\\\hline
\end{tabular}
^{T}\nonumber\\
&  \otimes%
\begin{tabular}
[c]{|c|c|c|}\hline
$\nu_{2}^{1/2}$ & $\cdots$ & $\nu_{m_{1/2}}^{1/2}$\\\hline
\end{tabular}
^{T}\overset{k}{\underset{j\neq l}{\otimes}}%
\begin{tabular}
[c]{|c|}\hline
$\mu_{1}^{j}$\\\hline
\end{tabular}
\underset{m_{j}-1-l_{j}}{\underbrace{%
\begin{tabular}
[c]{|l|l|l|}\hline
$T$ & $\cdots$ & $T$\\\hline
\end{tabular}
}}\underset{l_{j}}{\underbrace{%
\begin{tabular}
[c]{|l|l|l|}\hline
$L$ & $\cdots$ & $L$\\\hline
\end{tabular}
}}\overset{s}{\underset{r\neq1/2,l+1/2}{\otimes}}%
\begin{tabular}
[c]{|c|c|c|}\hline
$\nu_{1}^{r}$ & $\cdots$ & $\nu_{m_{r}}^{r}$\\\hline
\end{tabular}
^{T}\nonumber\\
&  +\sum_{l\geq1}l\left(  m_{l}-1-l_{l}\right)
\begin{tabular}
[c]{|c|}\hline
$\mu_{1}^{l}$\\\hline
\end{tabular}
\underset{m_{l}-2-l_{l}}{\underbrace{%
\begin{tabular}
[c]{|l|l|l|}\hline
$T$ & $\cdots$ & $T$\\\hline
\end{tabular}
}}\underset{l_{l}}{\underbrace{%
\begin{tabular}
[c]{|l|l|l|}\hline
$L$ & $\cdots$ & $L$\\\hline
\end{tabular}
}}\otimes%
\begin{tabular}
[c]{|c|c|c|c|}\hline
$T$ & $\nu_{1}^{l+1/2}$ & $\cdots$ & $\nu_{m_{l+1/2}}^{l+1/2}$\\\hline
\end{tabular}
^{T}\nonumber\\
&  \otimes%
\begin{tabular}
[c]{|c|c|c|}\hline
$\nu_{2}^{1/2}$ & $\cdots$ & $\nu_{m_{1/2}}^{1/2}$\\\hline
\end{tabular}
^{T}\overset{k}{\underset{j\neq l}{\otimes}}%
\begin{tabular}
[c]{|c|}\hline
$\mu_{1}^{j}$\\\hline
\end{tabular}
\underset{m_{j}-1-l_{j}}{\underbrace{%
\begin{tabular}
[c]{|l|l|l|}\hline
$T$ & $\cdots$ & $T$\\\hline
\end{tabular}
}}\underset{l_{j}}{\underbrace{%
\begin{tabular}
[c]{|l|l|l|}\hline
$L$ & $\cdots$ & $L$\\\hline
\end{tabular}
}}\overset{s}{\underset{r\neq1/2,l+1/2}{\otimes}}%
\begin{tabular}
[c]{|c|c|c|}\hline
$\nu_{1}^{r}$ & $\cdots$ & $\nu_{m_{r}}^{r}$\\\hline
\end{tabular}
^{T}\nonumber\\
&  +\sum_{l\geq1}ll_{l}%
\begin{tabular}
[c]{|c|}\hline
$\mu_{1}^{l}$\\\hline
\end{tabular}
\underset{m_{l}-1-l_{l}}{\underbrace{%
\begin{tabular}
[c]{|l|l|l|}\hline
$T$ & $\cdots$ & $T$\\\hline
\end{tabular}
}}\underset{l_{l}-1}{\underbrace{%
\begin{tabular}
[c]{|l|l|l|}\hline
$L$ & $\cdots$ & $L$\\\hline
\end{tabular}
}}\otimes%
\begin{tabular}
[c]{|c|c|c|c|}\hline
$L$ & $\nu_{1}^{l+1/2}$ & $\cdots$ & $\nu_{m_{l+1/2}}^{l+1/2}$\\\hline
\end{tabular}
^{T}\nonumber\\
&  \otimes%
\begin{tabular}
[c]{|c|c|c|}\hline
$\nu_{2}^{1/2}$ & $\cdots$ & $\nu_{m_{1/2}}^{1/2}$\\\hline
\end{tabular}
^{T}\overset{k}{\underset{j\neq l}{\otimes}}%
\begin{tabular}
[c]{|c|}\hline
$\mu_{1}^{j}$\\\hline
\end{tabular}
\underset{m_{j}-1-l_{j}}{\underbrace{%
\begin{tabular}
[c]{|l|l|l|}\hline
$T$ & $\cdots$ & $T$\\\hline
\end{tabular}
}}\underset{l_{j}}{\underbrace{%
\begin{tabular}
[c]{|l|l|l|}\hline
$L$ & $\cdots$ & $L$\\\hline
\end{tabular}
}}\overset{s}{\underset{r\neq1/2,l+1/2}{\otimes}}%
\begin{tabular}
[c]{|c|c|c|}\hline
$\nu_{1}^{r}$ & $\cdots$ & $\nu_{m_{r}}^{r}$\\\hline
\end{tabular}
^{T}\nonumber\\
&  +\sum_{i=2}^{m_{1/2}}%
\begin{tabular}
[c]{|c|c|}\hline
$\nu_{i}^{1/2}$ & $\mu_{1}^{1}$\\\hline
\end{tabular}
\underset{m_{1}-1-l_{1}}{\underbrace{%
\begin{tabular}
[c]{|l|l|l|}\hline
$T$ & $\cdots$ & $T$\\\hline
\end{tabular}
}}\underset{l_{1}}{\underbrace{%
\begin{tabular}
[c]{|l|l|l|}\hline
$L$ & $\cdots$ & $L$\\\hline
\end{tabular}
}}\otimes\left(  -1\right)  ^{i+1}%
\begin{tabular}
[c]{|c|c|c|c|c|}\hline
$\nu_{2}^{1/2}$ & $\cdots$ & $\hat{\nu}_{i}^{1/2}$ & $\cdots$ & $\nu_{m_{1/2}%
}^{1/2}$\\\hline
\end{tabular}
^{T}\nonumber\\
&  \overset{k}{\underset{j\neq1}{\otimes}}%
\begin{tabular}
[c]{|c|}\hline
$\mu_{1}^{j}$\\\hline
\end{tabular}
\underset{m_{j}-1-l_{j}}{\underbrace{%
\begin{tabular}
[c]{|l|l|l|}\hline
$T$ & $\cdots$ & $T$\\\hline
\end{tabular}
}}\underset{l_{j}}{\underbrace{%
\begin{tabular}
[c]{|l|l|l|}\hline
$L$ & $\cdots$ & $L$\\\hline
\end{tabular}
}}\overset{s}{\underset{r\neq1/2}{\otimes}}%
\begin{tabular}
[c]{|c|c|c|}\hline
$\nu_{1}^{r}$ & $\cdots$ & $\nu_{m_{r}}^{r}$\\\hline
\end{tabular}
^{T}\nonumber\\
&  +\sum_{l\geq2}\sum_{i=1}^{m_{l-1/2}}\left(  -1\right)  ^{i+1}%
\begin{tabular}
[c]{|c|c|}\hline
$\nu_{i}^{l-1/2}$ & $\mu_{1}^{l}$\\\hline
\end{tabular}
\underset{m_{l}-1-l_{l}}{\underbrace{%
\begin{tabular}
[c]{|l|l|l|}\hline
$T$ & $\cdots$ & $T$\\\hline
\end{tabular}
}}\underset{l_{l}}{\underbrace{%
\begin{tabular}
[c]{|l|l|l|}\hline
$L$ & $\cdots$ & $L$\\\hline
\end{tabular}
}}\otimes%
\begin{tabular}
[c]{|c|c|c|c|c|}\hline
$\nu_{1}^{l-1/2}$ & $\cdots$ & $\hat{\nu}_{i}^{l-1/2}$ & $\cdots$ &
$\nu_{m_{l-1/2}}^{l-1/2}$\\\hline
\end{tabular}
^{T}\nonumber\\
&  \otimes%
\begin{tabular}
[c]{|c|c|c|}\hline
$\nu_{2}^{1/2}$ & $\cdots$ & $\nu_{m_{1/2}}^{1/2}$\\\hline
\end{tabular}
^{T}\overset{k}{\underset{j\neq l}{\otimes}}%
\begin{tabular}
[c]{|c|}\hline
$\mu_{1}^{j}$\\\hline
\end{tabular}
\underset{m_{j}-1-l_{j}}{\underbrace{%
\begin{tabular}
[c]{|l|l|l|}\hline
$T$ & $\cdots$ & $T$\\\hline
\end{tabular}
}}\underset{l_{j}}{\underbrace{%
\begin{tabular}
[c]{|l|l|l|}\hline
$L$ & $\cdots$ & $L$\\\hline
\end{tabular}
}}\overset{s}{\underset{r\neq1/2,l-1/2}{\otimes}}%
\begin{tabular}
[c]{|c|c|c|}\hline
$\nu_{1}^{r}$ & $\cdots$ & $\nu_{m_{r}}^{r}$\\\hline
\end{tabular}
^{T},
\end{align}%
\begin{align}
0  &  =M%
\begin{tabular}
[c]{|c|c|c|c|}\hline
$L$ & $\nu_{2}^{3/2}$ & $\cdots$ & $\nu_{m_{3/2}}^{3/2}$\\\hline
\end{tabular}
^{T}\overset{k}{\underset{j=1}{\otimes}}%
\begin{tabular}
[c]{|c|}\hline
$\mu_{1}^{j}$\\\hline
\end{tabular}
\underset{m_{j}-1-l_{j}}{\underbrace{%
\begin{tabular}
[c]{|l|l|l|}\hline
$T$ & $\cdots$ & $T$\\\hline
\end{tabular}
}}\underset{l_{j}}{\underbrace{%
\begin{tabular}
[c]{|l|l|l|}\hline
$L$ & $\cdots$ & $L$\\\hline
\end{tabular}
}}\overset{s}{\underset{r\neq3/2}{\otimes}}%
\begin{tabular}
[c]{|c|c|c|}\hline
$\nu_{1}^{r}$ & $\cdots$ & $\nu_{m_{r}}^{r}$\\\hline
\end{tabular}
^{T}\nonumber\\
&  +%
\begin{tabular}
[c]{|c|}\hline
$\mu_{1}^{1}$\\\hline
\end{tabular}
\underset{m_{1}-l_{1}}{\underbrace{%
\begin{tabular}
[c]{|l|l|l|}\hline
$T$ & $\cdots$ & $T$\\\hline
\end{tabular}
}}\underset{l_{1}}{\underbrace{%
\begin{tabular}
[c]{|l|l|l|}\hline
$L$ & $\cdots$ & $L$\\\hline
\end{tabular}
}}\otimes%
\begin{tabular}
[c]{|c|c|c|c|}\hline
$T$ & $\nu_{1}^{1/2}$ & $\cdots$ & $\nu_{m_{1/2}}^{1/2}$\\\hline
\end{tabular}
^{T}\otimes%
\begin{tabular}
[c]{|c|c|c|}\hline
$\nu_{2}^{3/2}$ & $\cdots$ & $\nu_{m_{3/2}}^{3/2}$\\\hline
\end{tabular}
^{T}\nonumber\\
&  \overset{k}{\underset{j\neq1}{\otimes}}%
\begin{tabular}
[c]{|c|}\hline
$\mu_{1}^{j}$\\\hline
\end{tabular}
\underset{m_{j}-1-l_{j}}{\underbrace{%
\begin{tabular}
[c]{|l|l|l|}\hline
$T$ & $\cdots$ & $T$\\\hline
\end{tabular}
}}\underset{l_{j}}{\underbrace{%
\begin{tabular}
[c]{|l|l|l|}\hline
$L$ & $\cdots$ & $L$\\\hline
\end{tabular}
}}\overset{s}{\underset{r\neq1/2,3/2}{\otimes}}%
\begin{tabular}
[c]{|c|c|c|}\hline
$\nu_{1}^{r}$ & $\cdots$ & $\nu_{m_{r}}^{r}$\\\hline
\end{tabular}
^{T}\nonumber\\
&  +\sum_{l\geq1}l\underset{m_{l}-1-l_{l}}{\underbrace{%
\begin{tabular}
[c]{|l|l|l|}\hline
$T$ & $\cdots$ & $T$\\\hline
\end{tabular}
}}\underset{l_{l}}{\underbrace{%
\begin{tabular}
[c]{|l|l|l|}\hline
$L$ & $\cdots$ & $L$\\\hline
\end{tabular}
}}\otimes%
\begin{tabular}
[c]{|c|c|c|c|}\hline
$\mu_{1}^{l}$ & $\nu_{1}^{l+3/2}$ & $\cdots$ & $\nu_{m_{l+3/2}}^{l+3/2}%
$\\\hline
\end{tabular}
^{T}\nonumber\\
&  \otimes%
\begin{tabular}
[c]{|c|c|c|}\hline
$\nu_{2}^{3/2}$ & $\cdots$ & $\nu_{m_{3/2}}^{3/2}$\\\hline
\end{tabular}
^{T}\overset{k}{\underset{j\neq l}{\otimes}}%
\begin{tabular}
[c]{|c|}\hline
$\mu_{1}^{j}$\\\hline
\end{tabular}
\underset{m_{j}-1-l_{j}}{\underbrace{%
\begin{tabular}
[c]{|l|l|l|}\hline
$T$ & $\cdots$ & $T$\\\hline
\end{tabular}
}}\underset{l_{j}}{\underbrace{%
\begin{tabular}
[c]{|l|l|l|}\hline
$L$ & $\cdots$ & $L$\\\hline
\end{tabular}
}}\overset{s}{\underset{r\neq3/2,l+3/2}{\otimes}}%
\begin{tabular}
[c]{|c|c|c|}\hline
$\nu_{1}^{r}$ & $\cdots$ & $\nu_{m_{r}}^{r}$\\\hline
\end{tabular}
^{T}\nonumber\\
&  +\sum_{l\geq1}l\left(  m_{l}-1-l_{l}\right)
\begin{tabular}
[c]{|c|}\hline
$\mu_{1}^{l}$\\\hline
\end{tabular}
\underset{m_{l}-2-l_{l}}{\underbrace{%
\begin{tabular}
[c]{|l|l|l|}\hline
$T$ & $\cdots$ & $T$\\\hline
\end{tabular}
}}\underset{l_{l}}{\underbrace{%
\begin{tabular}
[c]{|l|l|l|}\hline
$L$ & $\cdots$ & $L$\\\hline
\end{tabular}
}}\otimes%
\begin{tabular}
[c]{|c|c|c|c|}\hline
$T$ & $\nu_{1}^{l+3/2}$ & $\cdots$ & $\nu_{m_{l+3/2}}^{l+3/2}$\\\hline
\end{tabular}
^{T}\nonumber\\
&  \otimes%
\begin{tabular}
[c]{|c|c|c|}\hline
$\nu_{2}^{3/2}$ & $\cdots$ & $\nu_{m_{3/2}}^{3/2}$\\\hline
\end{tabular}
^{T}\overset{k}{\underset{j\neq l}{\otimes}}%
\begin{tabular}
[c]{|c|}\hline
$\mu_{1}^{j}$\\\hline
\end{tabular}
\underset{m_{j}-1-l_{j}}{\underbrace{%
\begin{tabular}
[c]{|l|l|l|}\hline
$T$ & $\cdots$ & $T$\\\hline
\end{tabular}
}}\underset{l_{j}}{\underbrace{%
\begin{tabular}
[c]{|l|l|l|}\hline
$L$ & $\cdots$ & $L$\\\hline
\end{tabular}
}}\overset{s}{\underset{r\neq3/2,l+3/2}{\otimes}}%
\begin{tabular}
[c]{|c|c|c|}\hline
$\nu_{1}^{r}$ & $\cdots$ & $\nu_{m_{r}}^{r}$\\\hline
\end{tabular}
^{T}\nonumber\\
&  +\sum_{l\geq1}ll_{l}%
\begin{tabular}
[c]{|c|}\hline
$\mu_{1}^{l}$\\\hline
\end{tabular}
\underset{m_{l}-1-l_{l}}{\underbrace{%
\begin{tabular}
[c]{|l|l|l|}\hline
$T$ & $\cdots$ & $T$\\\hline
\end{tabular}
}}\underset{l_{l}-1}{\underbrace{%
\begin{tabular}
[c]{|l|l|l|}\hline
$L$ & $\cdots$ & $L$\\\hline
\end{tabular}
}}\otimes%
\begin{tabular}
[c]{|c|c|c|c|}\hline
$L$ & $\nu_{1}^{l+3/2}$ & $\cdots$ & $\nu_{m_{l+3/2}}^{l+3/2}$\\\hline
\end{tabular}
^{T}\nonumber\\
&  \otimes%
\begin{tabular}
[c]{|c|c|c|}\hline
$\nu_{2}^{3/2}$ & $\cdots$ & $\nu_{m_{3/2}}^{3/2}$\\\hline
\end{tabular}
^{T}\overset{k}{\underset{j\neq l}{\otimes}}%
\begin{tabular}
[c]{|c|}\hline
$\mu_{1}^{j}$\\\hline
\end{tabular}
\underset{m_{j}-1-l_{j}}{\underbrace{%
\begin{tabular}
[c]{|l|l|l|}\hline
$T$ & $\cdots$ & $T$\\\hline
\end{tabular}
}}\underset{l_{j}}{\underbrace{%
\begin{tabular}
[c]{|l|l|l|}\hline
$L$ & $\cdots$ & $L$\\\hline
\end{tabular}
}}\overset{s}{\underset{r\neq3/2,l+3/2}{\otimes}}%
\begin{tabular}
[c]{|c|c|c|}\hline
$\nu_{1}^{r}$ & $\cdots$ & $\nu_{m_{r}}^{r}$\\\hline
\end{tabular}
^{T}\nonumber\\
&  +\sum_{i=2}^{m_{3/2}}3%
\begin{tabular}
[c]{|c|c|}\hline
$\nu_{i}^{3/2}$ & $\mu_{1}^{3}$\\\hline
\end{tabular}
\underset{m_{3}-1-l_{3}}{\underbrace{%
\begin{tabular}
[c]{|l|l|l|}\hline
$T$ & $\cdots$ & $T$\\\hline
\end{tabular}
}}\underset{l_{3}}{\underbrace{%
\begin{tabular}
[c]{|l|l|l|}\hline
$L$ & $\cdots$ & $L$\\\hline
\end{tabular}
}}\otimes\left(  -1\right)  ^{i+1}%
\begin{tabular}
[c]{|c|c|c|c|c|}\hline
$\nu_{2}^{3/2}$ & $\cdots$ & $\hat{\nu}_{i}^{3/2}$ & $\cdots$ & $\nu_{m_{3/2}%
}^{3/2}$\\\hline
\end{tabular}
^{T}\nonumber\\
&  \overset{k}{\underset{j\neq3}{\otimes}}%
\begin{tabular}
[c]{|c|}\hline
$\mu_{1}^{j}$\\\hline
\end{tabular}
\underset{m_{j}-1-l_{j}}{\underbrace{%
\begin{tabular}
[c]{|l|l|l|}\hline
$T$ & $\cdots$ & $T$\\\hline
\end{tabular}
}}\underset{l_{j}}{\underbrace{%
\begin{tabular}
[c]{|l|l|l|}\hline
$L$ & $\cdots$ & $L$\\\hline
\end{tabular}
}}\overset{s}{\underset{r\neq3/2}{\otimes}}%
\begin{tabular}
[c]{|c|c|c|}\hline
$\nu_{1}^{r}$ & $\cdots$ & $\nu_{m_{r}}^{r}$\\\hline
\end{tabular}
^{T}\nonumber\\
&  +\sum_{l\geq2,l\neq3}\sum_{i=1}^{m_{l-3/2}}%
\begin{tabular}
[c]{|c|c|}\hline
$\nu_{i}^{l-3/2}$ & $\mu_{1}^{l}$\\\hline
\end{tabular}
\underset{m_{l}-1-l_{l}}{\underbrace{%
\begin{tabular}
[c]{|l|l|l|}\hline
$T$ & $\cdots$ & $T$\\\hline
\end{tabular}
}}\underset{l_{l}}{\underbrace{%
\begin{tabular}
[c]{|l|l|l|}\hline
$L$ & $\cdots$ & $L$\\\hline
\end{tabular}
}}\otimes\left(  -1\right)  ^{i+1}%
\begin{tabular}
[c]{|c|c|c|c|c|}\hline
$\nu_{1}^{l-3/2}$ & $\cdots$ & $\hat{\nu}_{i}^{l-3/2}$ & $\cdots$ &
$\nu_{m_{l-3/2}}^{l-3/2}$\\\hline
\end{tabular}
^{T}\nonumber\\
&  \otimes%
\begin{tabular}
[c]{|c|c|c|}\hline
$\nu_{2}^{3/2}$ & $\cdots$ & $\nu_{m_{3/2}}^{3/2}$\\\hline
\end{tabular}
^{T}\overset{k}{\underset{j\neq l}{\otimes}}%
\begin{tabular}
[c]{|c|}\hline
$\mu_{1}^{j}$\\\hline
\end{tabular}%
^{T}.
\end{align}%
\end{subequations}%
There are some undetermined parameters, which can be $L$ or $T$, in the above
equations. However, it is easy to see that both choice lead to the same
equations. Therefore, we will set all of them to be $T$ in the following.
Thus, the constrain equations become%
\begin{subequations}%
\begin{align}
0  &  =M%
\begin{tabular}
[c]{|c|c|c|c|}\hline
$L$ & $\nu_{2}^{1/2}$ & $\cdots$ & $\nu_{m_{1/2}}^{1/2}$\\\hline
\end{tabular}%
^{T}.
\end{align}%
\end{subequations}%
Next, we will deal with those antisymmetric indices$\left\{  \nu_{i}%
^{r}\right\}  $. In this case, there are much fewer possibilities which we can
chosen, i.e.
\begin{equation}%
%
^{T}. \label{b}%
\end{align}

Using the lemma (\ref{lamma}), the equations (\ref{a}) and (\ref{b}) reduce to%
\begin{subequations}%
\begin{align}
0  &  =M\underset{m_{1}-l_{1}}{\underbrace{%
%
.
\end{align}%
\end{subequations}%
From the first equation we have:

For $\nu_{2}^{1/2}=0$, $\nu_{3}^{1/2}=0$ and $\nu_{1}^{3/2}=0,$%
\begin{align}
0  &  =M\underset{m_{1}-l_{1}}{\underbrace{%
%
. \label{1-L0L}%
\end{align}

From the second equation we have:

For $\nu_{2}^{1/2}=0$, $\nu_{3}^{1/2}=0$ and $\nu_{1}^{3/2}=0,$%
\begin{align}
0  &  =M\underset{m_{1}-1-l_{1}}{\underbrace{%
%
. \label{3-L0L}%
\end{align}
Using the equations (\ref{1-000}) and (\ref{1-L0L}), we get%
\begin{align}
&  \underset{m_{1}-l_{1}}{\underbrace{%
%
,
\end{align}
then using the equations (\ref{1-T0L}), (\ref{3-000}) and (\ref{3-L0L}), we
obtain
\begin{align}
&  \underset{n-2m_{2}-2k}{\underbrace{%
%
.
\end{align}

\end{document}